\documentclass[fleqn,usenatbib]{mnras}

\usepackage{newtxtext,newtxmath}

\usepackage[T1]{fontenc}

\DeclareRobustCommand{\VAN}[3]{#2}
\let\VANthebibliography\thebibliography
\def\thebibliography{\DeclareRobustCommand{\VAN}[3]{##3}\VANthebibliography}

\usepackage{amsmath}	% Advanced maths commands
\usepackage{newtxtext,newtxmath}
\usepackage{pdfpages}
\usepackage{float}
\usepackage{tabularx}
\usepackage{comment}
\usepackage[normalem]{ulem}
\usepackage[mathlines, pagewise]{lineno}
\usepackage{colortbl}

\usepackage{subcaption}
\usepackage{graphicx}	% Including figure files
\usepackage{amsmath}	% Advanced maths commands
\usepackage{hyperref}
\usepackage{physics}
\usepackage{arydshln}
\usepackage{multirow}

\newcommand{\eightsigma}{$f_\mathrm{gas}~-8\sigma$}
\newcommand{\pk}{$P(k) / P_\mathrm{DM~Only}(k)$}
\title[Constraining baryon feedback]{The suppression of the matter power spectrum: strong feedback from X-ray gas mass fractions, kSZ effect profiles, and galaxy-galaxy lensing}

\author[J. Siegel et al.]{
\parbox{\textwidth}{
\Large Jared Siegel,$^{1}$\thanks{E-mail: siegeljc@princeton.edu}
Leah Bigwood,$^{2,3}$
Alexandra Amon,$^{1}$
Jamie McCullough,$^{1}$
Masaya Yamamoto,$^{1}$\\
Ian G. McCarthy,$^{4}$
Matthieu Schaller,$^{5,6}$
Aurel Schneider,$^{7}$
Joop Schaye$^{6}$\\
\small
$^{1}$Department of Astrophysical Sciences, Princeton University, 4 Ivy Lane, Princeton, NJ 08544, USA\\
$^{2}$ Institute of Astronomy and Kavli Institute for Cosmology, University of Cambridge, Madingley Road, Cambridge, CB3 0HA, UK\\
$^{3}$Kavli Institute for Cosmology (KICC), University of Cambridge, Madingley Road, Cambridge CB3 0HA, UK\\
$^{4}$Astrophysics Research Institute, Liverpool John Moores University, Liverpool, L3 5RF, UK\\
$^5$Lorentz Institute for Theoretical Physics, Leiden University, PO box 9506, 2300 RA Leiden, the Netherlands\\
$^6$Leiden Observatory, Leiden University, PO Box 9513, 2300 RA Leiden, the Netherlands\\
$^{7}$Department of Astrophysics, University of Zurich, Winterthurerstrasse 190, 8057 Zurich,
Switzerland\\}}

\date{Accepted XXX. Received YYY; in original form ZZZ}

\pubyear{\the\year{}}

\begin{document}
\label{firstpage}
\pagerange{\pageref{firstpage}--\pageref{lastpage}}
\maketitle

\begin{abstract}
Baryon feedback redistributes gas relative to the underlying dark matter distribution and suppresses the matter power spectrum on small scales, but the amplitude and scale dependence of this effect are uncertain.
We constrain the impact of baryon feedback on the matter power spectrum by jointly analysing X-ray gas mass fractions from the eROSITA and HSC--XXL samples and SDSS/DESI+ACT kinetic Sunyaev–Zel’dovich (kSZ) effect profiles; 
the samples are characterised with galaxy-galaxy lensing and together span group and cluster masses ($13 < \log_{10} \frac{M_{500}}{\mathrm{M}_\odot~h^{-1}} < 14$) at $0<z<1$. 
Using the baryonification framework, our joint eROSITA and kSZ model gives precise constraints on the suppression of the matter power spectrum: $10\pm2\%$ at $k=1~h~\mathrm{Mpc}^{-1}$.
The inferred gas profiles are more extended and the power suppression is stronger than predicted by the fiducial models of recent hydrodynamical simulation suites, including FLAMINGO and BAHAMAS.
The HSC--XXL gas mass fractions, which the fiducial simulations were calibrated to reproduce, prefer more moderate power suppression than the kSZ and eROSITA data: $5\pm4\%$ at $k=1~h~\mathrm{Mpc}^{-1}$.
With a simulated LSST Year~1 weak lensing analysis, we demonstrate a framework for next-generation surveys: calibrating feedback models with multi-wavelength gas observables to recover the small-scale statistical power of cosmic shear.
\end{abstract}


%%%%%%%%%%%%%%%%%%%%%%%%%%%%%%%%%%%%%%%%%%%%%%%%%%

%%%%%%%%%%%%%%%%% BODY OF PAPER %%%%%%%%%%%%%%%%%%

\section{Introduction}

Baryon feedback processes, particularly energy injection by active galactic nuclei (AGN), redistribute gas relative to the underlying dark matter distribution. This redistribution suppresses the matter power spectrum on small, non-linear scales by tens of percent relative to dark matter-only predictions \citep[e.g.,][]{vanDaalen2011,Chisari2019,Debackere2020}. 
Testing the cosmological model on small scales  requires accurate modelling of the non-linear matter power spectrum \citep[e.g.,][]{Semboloni2011}. 
However, the precise amplitude and scale dependence of matter power suppression due to baryon feedback remains unclear.
This uncertainty directly translates to a loss of cosmological precision.
Modelling of baryonic feedback and the intrinsic alignment of galaxies \citep[correlations between galaxy shapes not introduced by weak lensing, e.g.,][]{Chisari2025} limit cosmological precision by more than a factor of two for cosmic shear \citep{Amon2022}---the weak gravitational lensing of background galaxies by large-scale structure.
With recent analyses partially lifting the uncertainty from intrinsic alignments by limiting shear studies to blue galaxy populations that are found to weakly align \citep{McCullough2024}, baryonic feedback is now the limiting systematic for cosmic shear.

There is no consensus from modern hydrodynamical simulations on the impact of baryon feedback on the matter power spectrum.
Simulations include a variety of baryonic feedback processes (e.g., AGN, supernovae, and stellar winds) and are increasingly in agreement with the observed stellar-to-halo mass relation, cosmic star formation density, and cluster scaling relations  \citep[e.g.,][]{McCarthy2017, Dave2019, Schaye2023, Pakmor2023, Dolag2025, Schaye2025,Bigwood2025XFABLE}. 
However, they lack the resolution to treat feedback processes from first principles and instead adopt ``sub-grid'' prescriptions for the behaviour of stars and AGN.
Different sub-grid physics recipes result in markedly different predictions for the suppression of the matter power spectrum \citep[e.g.,][]{vanDaalen2011,vanDaalen2020,Bigwood2025XFABLE}.

Without a precise prediction from simulations for the impact of feedback on the matter power spectrum, weak lensing analyses often adopt conservative scale cuts: excluding the small-scale modes most sensitive to baryonic physics. 
Although it can reduce bias, this approach discards a large fraction of the available signal. 
An alternative method is to retain the small scales and account for matter power suppression by baryon feedback using flexible models: e.g., the halo model  \citep{Semboloni2011,Semboloni2013,Mead2021}, the baryonification of dark matter-only simulations \citep{Schneider2015, Angulo2021, Arico2021, Schneider2025}, Principal Component analysis \citep[PCA,][]{Huang2021}, the resummation model \citep{vanLoon2024,vanDaalen2025},  simulation-based analytic models \citep{Salcido2023,Schaller2025analytic},  or simulator based emulators \citep[e.g.,][]{Schaller2025}. 
Without calibration from external probes of baryon feedback (e.g., the hot gas mass fractions of groups and clusters), the flexibility of these models effectively downweights the small scales, degrading the cosmological precision similarly to scale cuts \citep{Arico2023, bigwood2024, anbajagane2025}.

In this work, we use multi-wavelength gas observations and galaxy-galaxy lensing to directly constrain the distribution of baryons relative to the underlying dark matter.
We then relate these constraints to the suppression of the matter power spectrum, producing data-driven priors that can be incorporated into weak lensing analyses. 
This approach transforms feedback modelling from a source of systematic uncertainty to an observationally calibrated component of cosmological inference.

By tracing how much gas is expelled from halos and how far it extends, observations of the intracluster medium (ICM) are valuable probes of baryon feedback. 
Recent studies have inferred the impact of baryon feedback on the matter power spectrum using a variety of observations, including X-ray gas mass fractions \citep{Schneider2022,bigwood2024,Grandis2024,Kovac25,vanDaalen2025}, diffuse X-ray emission \citep{ferreira2024,LaPosta2025}, the thermal and kinetic Sunyaev–Zel’dovich effects \citep[tSZ and kSZ,][]{Schneider2022,Troster2022,bigwood2024,Pandey2025,Dalal2025,Kovac25,Roper2025}, and fast radio bursts \citep{reischke2025}.
The matter power spectrum is indirectly benchmarked by comparing hydrodynamical simulations to gas observables \citep{McCarthy2018,Schaye2023,McCarthy2025, Hadzhiyska2024photoz, Ried2025, Siegel2025flamingo, Bigwood2025allthesims}.
The suppression of the non-linear matter power spectrum has also been constrained directly from the cosmic shear data, either with or without assuming the $\Lambda$CDM cosmology preferred by the CMB \citep[e.g.][]{Huang2021,Chen2022,AmonEfstathiou22, Preston23, Arico2023, GarciaGarcia2024, terasawa2024, Preston2024, sarmiento2025, broxterman2025,xu2025}.
Overall, there is a growing consensus that feedback indeed suppresses the matter power spectrum on small scales, and more recently there is evidence that suppression may be stronger than realised in many state-of-the-art hydrodynamical simulations.
In addition to the effects of feedback, deviations of the non-linear matter power from the $\Lambda$CDM expectation could be signs of new physics \citep[e.g.][]{Hooper2022, Poulin2023, Rogers2023, Preston2025}. 
Motivated by these previous studies, we address key uncertainties in the study of baryon feedback and deliver precise priors for next-generation cosmic shear analyses.

Cosmic shear is sensitive to baryon feedback over a wide range of halo mass and redshift \citep[e.g.,][]{LucieSmith2025}.
At $0.1 < k ~h~\mathrm{Mpc}^{-1} < 10$ and $0<z<1$, suppression of the matter power spectrum is strongly correlated with the baryon properties of groups and clusters \citep{Salcido2023}. 
However, no single observation probes the full regime: X-ray gas mass fractions probe the inner regions ($<R_{500}$) of low redshift clusters, while existing kSZ measurements constrain the outskirts (beyond several $R_{500}$) of higher redshift group mass halos.
Building a complete picture of baryon feedback thus requires a multi-probe approach.
In this work, we jointly analyse stacked measurements of the kSZ effect from SDSS/DESI$+$ACT \citep{Schaan2021, Ried2025} and X-ray gas mass fractions from the first eROSITA All-Sky Survey catalogue \citep[RASS1,][]{Bulbul2024, Kluge2024} and the HSC--XXL sample \citep{Pierre2016,Eckert2016,Akino2022}.
For the kSZ and eROSITA samples, we adopt the galaxy-galaxy lensing mean halo mass measurements of \cite{Siegel2025flamingo}; the HSC--XXL sample includes the weak lensing halo mass measurements of \cite{Umetsu2020}.
Weak lensing halo mass constraints resolve a key source of uncertainty, because the implied magnitude of gas expulsion by a kSZ or X-ray observation is degenerate with the halo mass \citep[e.g.,][]{McCarthy2025}.

In Section~\ref{sec:data}, we describe the multi-wavelength datasets.
We review the baryonification model in Section~\ref{sec:bfc} and  validate the framework in Section~\ref{sec:inj}.
We present the model fits in Section~\ref{sec:results}.
In Section~\ref{sec:disc}, we compare our results with state-of-the-art hydrodynamical simulations and demonstrate how the model constraints can be implemented as data-driven priors for cosmic shear, using a forecast of LSST Year~1. 
We conclude in Section~\ref{sec:conc}.

\section{Data}
\label{sec:data}

In this work, we infer the impact of baryons on the matter power spectrum by jointly analysing kSZ effect profiles and X-ray hot gas mass fractions, with mean halo mass constraints from GGL.  
We describe the kSZ and X-ray measurements in Sections~\ref{sec:ksz} and \ref{sec:xray}, respectively. 

\subsection{SDSS/DESI+ACT kSZ effect profiles}
\label{sec:ksz}

The kinetic Sunyaev-Zel’dovich (kSZ) effect is a CMB secondary anisotropy, arising from the inverse Compton scattering of CMB photons off the ionised gas around galaxies and clusters.
Due to the bulk motion of the gas, these scatterings impart a Doppler shift to the CMB photons \citep[][]{Sunyaev1980}.
The apparent change in temperature induced by the kSZ effect $\Delta T_{\mathrm{kSZ}}$ is therefore proportional to the electron number density and the peculiar velocity of the gas:
\begin{equation}\label{eq:ksz1}
    \frac{\Delta T_{\mathrm{kSZ}}}{T_{\mathrm{CMB}}}=\frac{\sigma_{\rm T}}{c} \int_{\rm los} e^{-\tau} n_{\rm e} v_{\rm p} dl\, ,
\end{equation}
where $\sigma_{\rm T}$ is the Thomson scattering cross-section, $c$ is the speed of light, $n_\mathrm{e}$ is the free-electron number density, $v_{\rm p}$ is the peculiar velocity along the line of sight, and $\tau = \sigma_\mathrm{T}  \int n_\mathrm{e}  d l$ is the optical depth to Thompson scattering.
Because the gas of interest is optically thin, we assume $e^{-\tau}\approx1$ \citep[see][]{Schaan2021,Ried2025}.
For one galaxy, the kSZ effect is approximately
\begin{align}
    \frac{ \Delta T_\mathrm{kSZ} }{ T_\mathrm{CMB} } \approx - \tau_\mathrm{gal} \frac{ v_{ \mathrm{p}, \mathrm{gal}}}{ c }.
\end{align}

To achieve a high signal-to-noise ratio measurement, the kSZ signal must be stacked across many galaxies.  
We consider the kSZ effect profile measurements of \cite{Schaan2021} and \cite{Ried2025}, which employ compensated aperture
photometry (CAP) filters.
For a galaxy located at $\vec{\theta}$ on the sky, the CAP filter sums the temperature fluctuation within $\theta_d$ of the galaxy and background subtracts the temperature fluctuations between $\theta_d$ and $\sqrt{2} \theta_d$ \citep[Equations 8 and 9 of][]{Schaan2021}.

Because the kSZ effect is proportional to the velocity of the gas, the stacked kSZ signal is measured by a combination of velocity and inverse-variance weighting \citep{Schaan2021,Ried2025}:
\begin{equation}
    \label{eqn:ksz_stack}
    \hat{T}_\mathrm{kSZ}(\theta_d) = - \frac{1}{r_{v, \mathrm{bias}}} \frac{ v^\mathrm{rms}_\mathrm{rec} }{c} \frac{ \sum_i \mathcal{T}_i(\theta_d) ( \frac{v_{i, \mathrm{rec}}}{c}) \sigma^{-2}_i }{ \sum_i ( \frac{v_{i, \mathrm{rec}}}{c})^2 \sigma^{-2}_i },
\end{equation}
where $\mathcal{T}_i(\theta_d)$ is the CAP filtered signal for the $i$th galaxy, $\sigma_i^2$ is the variance in $\mathcal{T}_i(\theta_d)$, $v_{i, \mathrm{rec}}$ is the reconstructed line of sight peculiar velocity, and $v^\mathrm{rms}_\mathrm{rec}$ is the standard deviation of the reconstructed peculiar velocities.
For existing measurements, the galaxies' three-dimensional peculiar velocities were reconstructed from the galaxy number overdensity field by the linearized continuity equation \citep{Padmanabhan2012,VargasMagana2017}.
To account for biases in the peculiar velocity reconstruction, \cite{Schaan2021} and \cite{Ried2025} scale the profile amplitudes by the correlation between the true and reconstructed velocities in mock catalogues:
\begin{equation}
    \label{eqn:rvbias}
    r_{v, \mathrm{bias}} = \frac{ \langle v_\mathrm{true} v_\mathrm{rec} \rangle }{ v_\mathrm{true}^\mathrm{rms} v_\mathrm{rec}^\mathrm{rms} }.
\end{equation}
Using the \textsc{AbacusSummit} dark matter-only N-body simulation suite, \cite{Hadzhiyska2024} report the reconstruction quality for DESI-like spectroscopic samples: $r_{v, \mathrm{bias}} \sim 0.7$ with $<5\%$ uncertainties.
In Appendix~\ref{app:two_halo}, we compare the corrected line-of-sight RMS of the reconstructed velocities with matching samples from the FLAMINGO simulations, finding agreement within $10\%$.
For our modelling, we assume a conservative uncertainty of  $10\%$ for $r_{v, \mathrm{bias}}$.

In this work, we consider the stacked kSZ effect profiles of \cite{Schaan2021} and \cite{Ried2025}, which used spectroscopic redshift tracers from SDSS and DESI, respectively.
The ACT+Planck CMB temperature maps and galaxy samples underlying these measurements are described in Sections~\ref{sec:ksz_BOSS} and \ref{sec:ksz_DESI}, respectively.
Because the kSZ stacks are constructed with common CMB temperature maps (ACT~DR5 for SDSS and DR6 for DESI), the different measurements  (e.g. SDSS LOWZ and CMASS) are formally covariant.
The SDSS and DESI spectroscopic samples also share common objects \citep{desidr1}. 
For our analysis, we assume the kSZ measurements are independent; the galaxy selections occupy distinct comoving volumes and mass/redshift regimes, so map-level covariance contributes negligibly to the parameter uncertainties relative to kSZ measurement noise.
Investigation of measurement covariances with forward modelling is warranted.

\subsubsection{SDSS BOSS: LOWZ and CMASS}
\label{sec:ksz_BOSS}

\cite{Schaan2021} measured the stacked kSZ effect profiles for the LOWZ and CMASS samples from the Baryon Oscillation Spectroscopic Survey (BOSS) DR10 release \citep{Ahn2014}.
The LOWZ (``low redshift'') sample consists of red galaxies at $0.15 < z < 0.4$, and the CMASS (``constant mass'') sample includes massive galaxies at $0.4 < z < 0.8$ \citep{Ahn2012}. 
The CMB temperature map is derived from ACT DR5 \citep[][]{Fowler2007, Swetz2011,Thornton2016,Henderson2016} and Planck \citep{Planck2020}.
\cite{Schaan2021} measured the kSZ effect in both the 90 and 150~GHz frequency ACT maps;
we consider both measurements.
From the mock BOSS catalogues \citep{Manera2013, Manera2015}, \cite{Schaan2021} report a peculiar velocity bias factor of $r_{\rm v}=0.7$ and quote uncertainties of a few percent.

From \cite{Siegel2025flamingo}, we adopt GGL derived mean halo masses of $\log_{10}M_{500}/\mathrm{M}_\odot=13.51 \pm 0.01$ and $13.29 \pm 0.02$ for LOWZ and CMASS, respectively.

\subsubsection{DESI: BGS and LRG}
\label{sec:ksz_DESI}

\cite{Ried2025} measured the stacked kSZ effect profiles for two DESI~DR1 samples: the Bright Galaxy Survey \citep[BGS,][]{Hahn2023BGS} and the Luminous Red Galaxy sample \citep[LRG,][]{Zhou2023LRG}.
The BGS sample is magnitude limited and spans $0 < z < 0.6$.
Following \cite{DESItwopoint2024}, \cite{Ried2025} restricted the sample to $0.1 < z < 0.4$ and applied a luminosity selection as a function of redshift to create a sample with an approximately constant number density.
The LRG sample includes massive quenched galaxies at $0.4 < z < 1.1$.
Following \cite{Siegel2025flamingo}, we consider two of the stellar mass bins defined by \cite{Ried2025}: $\log_{10} M_*/\mathrm{M}_\odot$=$(10.5, 11.2), (11.2, 11.4)$; the mass selection is performed using the stellar masses from \cite{Zhou2023LRG}, which are derived from DESI Legacy Imaging photometry via a random forest algorithm.
\cite{Ried2025} adopt the harmonic-space Internal Linear Combination (hILC) CMB temperature map, which combines multiple ACT~DR6 channels ($90, 150,$ and $220$~GHz) with Planck \citep{Coulton2024}; each channel is convolved with a Gaussian beam ($1.6'$ FWHM) before the combination.
To mitigate tSZ contamination, \cite{Ried2025} masked temperature outliers in the ACT map.
Peculiar velocities were adopted from \cite{DESIDR1velocities2025}, with reconstruction biases of $r_{\rm v} = 0.64$ and $0.65$ for the BGS and LRG samples, respectively \citep{Hadzhiyska2024}; the reported $r_{\rm v}$ uncertainties are approximately $3\%$.

From \cite{Siegel2025flamingo}, we adopt GGL derived mean halo masses of $\log_{10}M_{500}/\mathrm{M}_\odot=13.31 \pm 0.01$,  $12.91 \pm 0.02$, and $13.15 \pm 0.01$ for BGS, LRG M1, and M2, respectively.

\subsection{X-ray gas mass fractions}
\label{sec:xray}

X-ray observations of the ICM in groups and clusters are a well established tracer of baryon feedback.
Stronger feedback processes are expected to expel gas beyond $R_{500}$, thereby lowering the typical gas fraction at a given halo mass. 
Indeed, the baryon fraction is found to increase as a function of radius and halo mass, approaching the universal baryon fraction at large radii of massive clusters \citep{Sun2009, Bulbul2012, Sanderson2013, Lovisari2015, Eckert2016, Eckert2019}.

We consider X-ray gas mass fraction measurements from the XMM-Newton XXL sample (Section~\ref{sec:xxl}) and the eROSITA All-Sky Survey (Section~\ref{sec:erosita}).
Weak lensing halo masses are available for both samples: \cite{Umetsu2020} and \cite{Siegel2025flamingo} for the XXL and eROSITA samples, respectively.
The GGL derived halo masses resolve a long standing uncertainty in the study of X-ray gas mass fractions. 
Previous studies often assumed hydrostatic equilibrium to estimate total masses, however,  
non-thermal pressure support can bias hydrostatic masses low by approximately $30\%$ \citep{Hoekstra2015,Eckert2016,Kugel2023,MunozEcheverria2024}; 
further complicating matters, this bias is dependent on cluster mass \citep{Braspenning2025}.

\subsubsection{HSC--XXL}
\label{sec:xxl}

The XXL survey targeted two $25$~degree$^2$ patches of the sky with over $6$~Ms of XMM-Newton observations \citep{Pierre2016}.
The second XXL data release includes over $300$ spectroscopically confirmed X-ray detected clusters at $0<z<1$ \citep{Adami2018}.
For the approximately $100$ clusters that overlap the HSC footprint, \cite{Umetsu2020} measured weak lensing masses with the first year HSC shear catalogue \citep{Mandelbaum2018}.
Using the weak lensing masses, \cite{Akino2022} measured the gas mass within $R_{500}$ for each XXL cluster in the HSC footprint and conducted a joint Bayesian analysis of the halo mass, gas mass, stellar mass, and X-ray luminosity scaling relations.
The X-ray selection effects were modelled as a minimum luminosity selection function \citep[Appendix~A1 of][]{Akino2022}.

With weak lensing halo masses and modelling of selection effects, the HSC--XXL scaling relations are particularly well characterised. 
The gas fraction scaling relation is broadly consistent with other pre-eROSITA X-ray gas fraction measurements \citep{Kugel2023}.
Following \cite{Kugel2023}, we approximate the HSC--XXL scaling relation as two halo mass bins:  $f_\mathrm{gas}(10^{13.5}~\mathrm{M}_\odot)= 0.054 \pm 0.010$
and $f_\mathrm{gas}(10^{14.5}~\mathrm{M}_\odot)= 0.106 \pm 0.02$.

\subsubsection{eROSITA All-Sky Survey}
\label{sec:erosita}

The eROSITA satellite \citep[extended ROentgen Survey with an Imaging Telescope Array,][]{Predehl2021} is a soft X-ray telescope on board the Spectrum-Roentgen-Gamma observatory  \citep[SRG,][]{Sunyaev2021}.  
The first eROSITA All-Sky Survey catalogue \citep[eRASS1,][]{Bulbul2024} includes more than $12,000$ optically confirmed X-ray groups and clusters in the Western Galactic Hemisphere between the local Universe and $z < 1.5$ \citep{Bulbul2024, Kluge2024}. 
The publicly available gas masses are measured by a parametric model fit to the X-ray images with a forward modelling calibration \citep{ Sanders2018,Liu2022, Bulbul2024}. 
The eRASS1 reported halo masses $M_{500}$ are estimated from a weak lensing calibrated scaling relation between redshift, X-ray count rate, and shear.
The shear measurements include corrections for miscentering and cluster member contamination, as well as calibration from forward modelling of simulated shear profiles \citep{Grandis2024, Kleinebreil2025, Okabe2025}. 

Although eROSITA is more complete than previous X-ray selected samples \citep[e.g., ROSAT,][]{Truemper1982,Voges1999}, the sample is still shaped by selection effects. 
The completeness of the eRASS1 catalogue significantly falls at $M_{500} \lesssim 10^{14}~\mathrm{M}_{\odot}$ \citep{Seppi2022,Popesso2024a,Marini2024}.
We therefore limit our study to lower redshift ($z<0.2$) massive halos.
Following \cite{Siegel2025flamingo},
we define four bins of eROSITA clusters: $\log_{10}M_{500}/\mathrm{M}_\odot=(13.3, 14.0)$ and $(14.0, 14.5)$ at $0.05 < z < 0.1$, and $\log_{10}M_{500}/\mathrm{M}_\odot=(13.5, 14.0)$ and $(14.0, 14.5)$ at $0.1 < z < 0.2$.
The four bins include a total of $2,500$ clusters.
\cite{Siegel2025flamingo} independently measured the mean halo masses for each bin with GGL, validating the publicly reported eRASS1 masses.
The GGL calibrated bins are consistent with the stacked X-ray gas mass fractions of optically selected halos \citep{Popesso2024}.

The eROSITA reported gas mass fractions suggest that groups and clusters are, on average, more gas-depleted than previously found  \citep{Popesso2024,Dev2024,Kovac25}.
At $M_{500} = 10^{14}~\mathrm{M}_\odot$, the mean gas mass from the eRASS1 catalogue is approximately $40\%$ lower than the HSC--XXL scaling relation.
We discuss potential explanations, including selection effects and modelling systematics, and avenues for further study in Section~\ref{sec:xray_disc}.
Because the apparent discrepancy in the mean gas fractions is not fully understood, we consider both the eROSITA and pre-eROSITA measurements in this work.

\section{Baryonification Model}
\label{sec:bfc}

\begin{figure*}
\centering
\includegraphics[width=\textwidth]{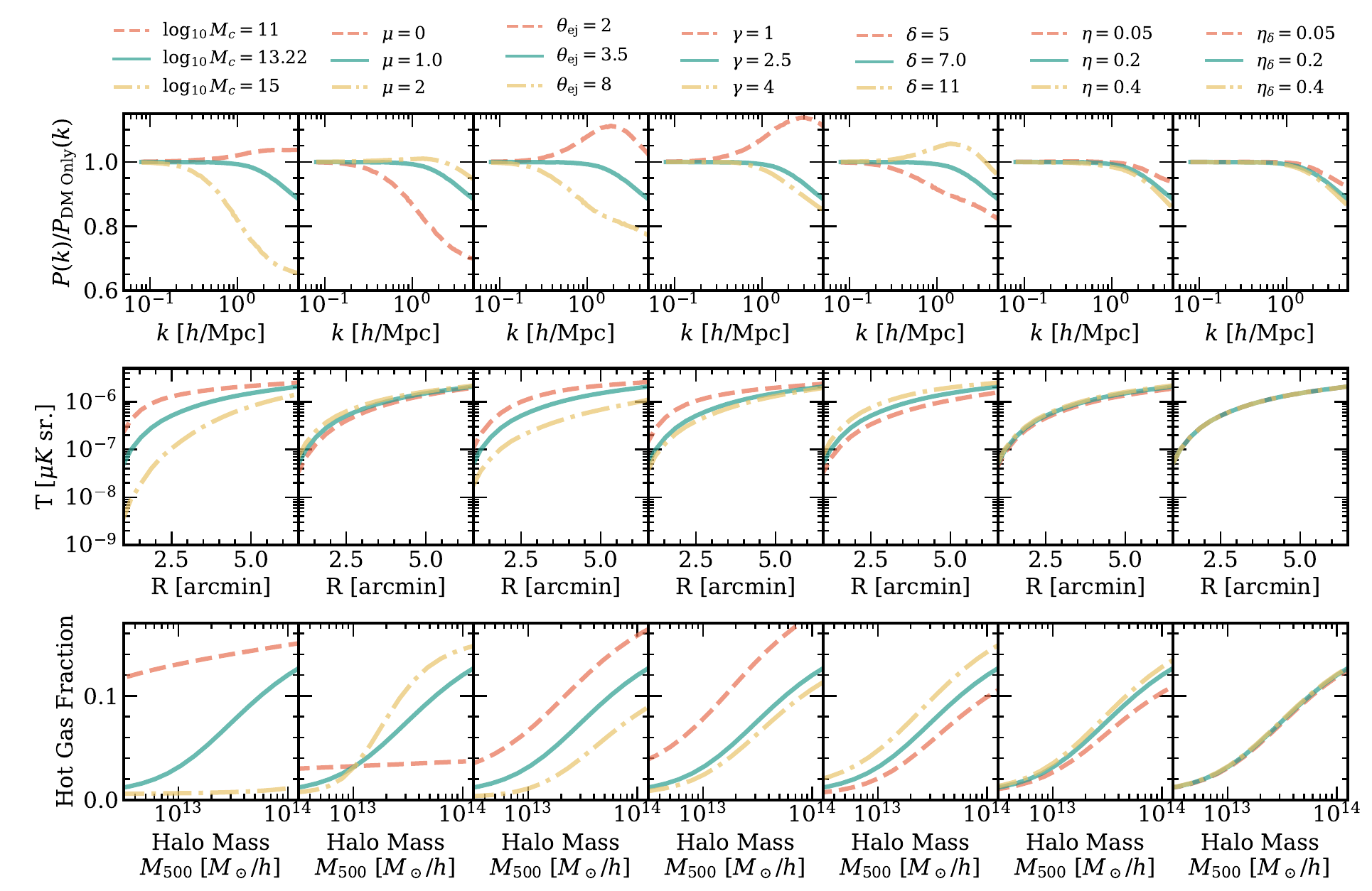}
    \caption{\label{fig:model} 
    The impact of the baryonification model parameters on the suppression of the matter power spectrum relative to a dark matter universe, $P(k)/P_{\rm DM~only}(k)$ (\textit{top}), the stacked kSZ effect profile as a function of angular radius for a $\log_{10}[M_{500}/\mathrm{M}_\odot~h^{-1}]=13.1$ halo at $z=0.3$  (\textit{middle}), and the hot gas mass fraction within $R_{500}$ as a function of halo mass (\textit{bottom}). 
    Each panel varies one baryonification parameter, within the prior bounds (Table~\ref{tab:priors}).
    For the leftmost column, $M_{\rm c}$ is reported in units of $\mathrm{M}_\odot~h^{-1}$.
    }
\end{figure*}

The ``baryonification" model is a framework for relating the radial density profiles of gas, stars, and dark matter to the suppression of the matter power spectrum relative to a dark matter-only simulation. 
The method relies on perturbatively displacing particles in N-body simulations according to observationally motivated, spherically symmetric analytic density profiles, to mimic the impact of feedback on the matter distribution \citep{Schneider2015,Giri2021,Schneider2025}.  
By calculating the power spectrum of the baryonified matter distribution, the suppression of the matter power spectrum can be related to the radial density profiles of gas, stars, and dark matter.
Observational constraints on the gas distribution from X-ray and kSZ measurements can thus be transformed into constraints on the matter power spectrum \citep{Schneider2022,Giri2021,Arico2023,bigwood2024,Kovac25}.
Power suppression can also be inferred with a halo model \citep{Debackere2020,LaPosta2025,Pandey2025godmax,Dalal2025}, see Section~\ref{sec:other_feedback_constraints}.

We outline the baryonification framework below and refer the reader to \cite{Schneider2015} and \cite{Giri2021} for a detailed description. 
The baryonification particle displacements convert Navarro–Frenk–White (NFW)-like profiles into baryonified halos composed of hot gas $\rho_{\rm gas}(r)$, a central galaxy $\rho_{\rm cga}(r)$, and collisionless matter $\rho_{\rm clm}(r)$:
\begin{equation}
\label{eq:bfc_profile}
\rho_{\rm nfw}(r)\,\,\rightarrow\,\,\rho_{\rm bfc}(r)= \rho_{\rm gas}(r) + \rho_{\rm cga}(r) + \rho_{\rm clm}(r) \,.
\end{equation}

The gas component is parametrised as a cored double-power law:
\begin{equation}
\label{eq:rhogas}
\rho_{\rm gas}(r) \propto \frac{\Omega_{\rm b}/\Omega_{\rm m}-f_{\rm star}(M_{200})}{\left[1+\left(\frac{r}{r_{\rm core}}\right)\right]^{\beta(M_{200})}\left[1+\left(\frac{r}{r_{\rm ej}}\right)^{\gamma}\right]^{\frac{\delta-\beta(M_{200})}{\gamma}}} \, ,
\end{equation}
where $\Omega_\mathrm{b} / \Omega_\mathrm{m}$ is the cosmic baryon fraction.
Inside the core radius $r_\mathrm{core} = 0.1 R_{200}$, the slope of the density profile is determined by the power law index
\begin{equation}
    \label{eq:beta}
    \beta(M_{200}) = \frac{3(M_{200} / M_\mathrm{c})^\mu}{1 +(M_{200} / M_\mathrm{c})^\mu}.
\end{equation}
At group masses ($M_{200} \ll M_\mathrm{c}$), the gas is efficiently expelled ($\beta \approx 0$), while for clusters ($M_{200} \gg M_\mathrm{c}$), the gas is retained ($\beta \approx 3$); the transition between the low and high mass limits is set by $\mu$.
Beyond the ejection radius ($r_{\rm ej}=\theta_{\rm ej}R_{200}$), the slope of the density profile is dictated by the parameters $\gamma$ and $\delta$.
The stellar fraction $f_\mathrm{sga}$ is parametrised as a power law $f_\mathrm{sga}(M_{200})=0.55 (M_{200} / M_s)^{-\eta}$, where $M_s=2.5 \times 10^{11}~\mathrm{M}_\odot~h^{-1}$  \citep{Moster2013}. 

The central galaxy component is parametrised as an exponentially truncated power
law
\begin{equation}
    \label{eq:cga}
    \rho_\mathrm{cga}(r) = \frac{ f_\mathrm{cga}(M_{200}) }{ 4 \pi^{3/2} R_\mathrm{h} r^2 } \exp \left[ - \left( \frac{r}{2R_\mathrm{h}} \right)^2 \right ],
\end{equation}
where $R_\mathrm{h}$ is the stellar half-light radius (fixed to $0.015$ the virial radius) and $f_\mathrm{cga}(M_{200})$ is the fraction of stars in the central galaxy: $f_\mathrm{cga}(M_{200})=0.55 (M_{200} / M_s)^{-(\eta+\eta_\delta)}$

The collisionless matter component, including dark matter, satellites, and halo stars, is parametrised as
\begin{equation}
    \label{eq:clm}
    \rho_\mathrm{clm}(r) = \left [ \frac{\Omega_\mathrm{dm}}{\Omega_\mathrm{m}} + f_\mathrm{sga}(M_{200}) \right ] \rho_{\rm nfw}(r),
\end{equation}
where $\Omega_\mathrm{dm} / \Omega_\mathrm{m}$ is the cosmic dark matter fraction and $f_\mathrm{sga} = f_\mathrm{star} - f_\mathrm{cga}$ is the stellar fraction in satellite galaxies and halo stars.

The seven free parameters $\mathbf{\Theta} = \{M_{\rm c}, \mu, \theta_{\rm ej}, \delta, \gamma, \eta, \eta_\delta\}$ are summarized in Table~\ref{tab:priors}.
We show how each parameter affects the suppression of the matter power spectrum, the kSZ signal, and the hot gas mass fractions of groups and clusters in Figure~\ref{fig:model}.
In this parametrisation of the baryonification model, the feedback parameters are redshift independent.
For a given set of model parameters $\mathbf{\Theta}$, the comoving density profile of each component only depends on halo mass and radius.
However, predictions of the baryonification model (e.g., gas mass fractions, kSZ effect profiles, or suppression of the matter power spectrum) still depend on redshift, because the clustering and scale factor of the underlying N-body simulation evolves with redshift.

In this work, we constrain the free parameters using GGL characterized X-ray and kSZ observations.
Stellar fraction measurements are not readily available for our bins of eROSITA clusters.
Measurements are available for other cluster samples \citep[e.g.,][]{Vikhlinin2006,Gonzalez2013, Sanderson2013,Lovisari2015,Kravtsov2018, Akino2022}, however, it is nontrivial to account for differences in sample selection and halo mass estimation between these studies and our eROSITA bins.
To ensure internal consistency, we refrain from modelling the stellar fraction for either the eROSITA or HSC--XXL X-ray samples. 
Following \cite{Kovac25}, we fix the stellar parameters to $\eta=0.1$ and $\eta_\delta=0.2$ for our fiducial analysis \citep[see][]{Moster2013}.
In Appendix~\ref{app:sensitivity}, we show that the parameter constraints are weakly sensitive to the adopted stellar parameters.
We repeat the fitting process for four different stellar parameter treatments, including conservative Gaussian priors instead of fixed values, and demonstrate that at the current statistical uncertainty, the effect of varying stellar priors is negligible.
We adopt wide top hat priors for the other five baryonification parameters. 
For our fiducial analysis, we fix the baryon fraction to the maximum likelihood cosmological parameters of the DES~Y3 `3×2pt + All Ext.' flat $\Lambda$CDM cosmology \citep{DES3x2}; we show our results are insensitive to this choice in Appendix~\ref{app:sensitivity}.

While the baryonification of a dark matter-only simulation is significantly faster than evolving a full hydrodynamical simulation, displacing particles in N-body simulations is still computationally expensive.
We therefore rely on the \texttt{BCemu} emulator \citep{Giri2021}, trained on $2,700$ baryonified simulations, for our Bayesian inference.
The underlying N-body simulation consisted of $512^3$ particles in a $256~(\mathrm{Mpc}~h^{-1})^3$ box with a $\Lambda$CDM cosmology: $\Omega_\mathrm{m}=0.315, \Omega_b=0.049, h_0 = 0.674, n_s = 0.96, \sigma_8 = 0.811$. 
Along with the seven radial profile parameters, the baryon fraction $\Omega_\mathrm{b} / \Omega_\mathrm{m}$ was varied between $0.10-0.25$.
\texttt{BCemu} has been demonstrated to be percent level accurate at $10^{-1}<k~[h~\mathrm{Mpc}~^{-1}]<10^{1}$ and $0<z<2$ \citep{Giri2021}.
A baryonification emulator is also publicly available from the BACCO simulation project \citep{Angulo2021, Arico2021}.
\cite{Schneider2025} recently introduced an updated version of the baryonification framework, including modified formulations of the density profiles and treatment of the dark matter, gas, and stellar components as independent fields; an emulator was not publicly available at the time of writing. 
In this work, we consider the original baryonification framework and reserve investigation of alternate formulations for future work (Shavelle et al. in prep).

\begin{table*}
\setlength\extrarowheight{3pt}
\begin{tabular}{ccccccc}
\hline
Parameter & Description & Prior & & Posterior & \\
 &  &  &  $f_\mathrm{gas}$ &  $f_\mathrm{gas}$ & kSZ & Joint kSZ \& \\
  &  &  &  HSC--XXL  & eROSITA &  & eROSITA $f_\mathrm{gas}$\\
\hline
$\log_{10} \frac{M_{\rm c}}{\mathrm{M}_\odot~h^{-1}}$ & The
characteristic mass scale at which the slope of the & U[11.0, 15.0] & $12.8_{-0.8}^{+0.6}$ &  $13.3_{-0.8}^{+0.8}$ &  $13.8_{-0.5}^{+0.6}$ &  $13.6_{-0.2}^{+0.2}$ 
 \\  & gas profile becomes shallower (Equation~\ref{eq:beta}). \\
$\theta_{\rm ej}$ & Specifies the characteristic radius of gas ejection:
 & U[2.0, 8.0] & $5.0_{-2.0}^{+2.0}$ &  $5.0_{-2.0}^{+2.0}$ &  $6.0_{-1.0}^{+1.0}$ &  $6.0_{-1.0}^{+1.0}$   \\
& $r_{\rm ej}=\theta_{\rm ej}R_{200}$ (Equation~\ref{eq:rhogas}). \\
$\mu$ & Defines
how fast the slope of the gas profile becomes & U[0.0, 2.0]& $0.5_{-0.3}^{+0.6}$ &  $0.5_{-0.2}^{+0.3}$ &  $0.4_{-0.3}^{+0.6}$ &  $0.9_{-0.1}^{+0.2}$  \\
& shallower towards small halo masses  (Equation~\ref{eq:beta}).\\ 
$\gamma$ & Exponent in gas profile parametrisation (Equation~\ref{eq:rhogas}) & U[1.0, 4.0]& $3.0_{-1.0}^{+1.0}$ &  $2.6_{-1.0}^{+0.9}$ &  $2.0_{-1.0}^{+1.0}$ &  $2.0_{-1.0}^{+1.0}$  \\
$\delta$ & Exponent in the gas profile parametrisation (Equation~\ref{eq:rhogas}) & U[5.0, 11.0]& $8.0_{-2.0}^{+2.0}$ &  $8.0_{-2.0}^{+2.0}$ &  $7.0_{-2.0}^{+2.0}$ &  $8.0_{-2.0}^{+2.0}$  \\
%\hdashline
$\eta$ & Specifies the total stellar fraction within a halo (Equation~\ref{eq:cga}) & Fixed: 0.1 &  \\
$\eta_{\delta}$ & Specifies the stellar fraction of the central galaxy:  & Fixed: 0.2 &  \\ 
&  $\eta_{\delta}= \eta_{\rm{cga}}-\eta$ (Equation~\ref{eq:cga}) \\
\hline
\end{tabular}
\caption{Qualitative descriptions of the model parameters, assumed priors, and posteriors from fitting the kSZ and X-ray data. 
We exclude the posteriors from the joint kSZ and HSC--XXL fit because of the poor goodness of fit; the posteriors are presented in Appendix~\ref{app:xray}. }
\label{tab:priors}
\end{table*}

\subsection{Modelling X-ray and kSZ measurements}

We constrain the free parameters of the baryonification model by jointly modelling the X-ray gas mass measurements (eROSITA or HSC--XXL) and the SDSS/DESI+ACT kSZ effect profiles.

At a given halo mass, we predict the X-ray gas mass by integrating the gas component $\rho_{\mathrm{gas}}(r,M)$ to $R_{500}$.
Previous studies often assumed hydrostatic equilibrium to derive cluster halo masses and therefore included a hydrostatic equilibrium bias  parameter \citep[e.g.,][]{Schneider2022,Kovac25}; we bypass this uncertainty by considering GGL halo masses: \cite{Umetsu2020}  and \cite{Siegel2025flamingo} for the HSC--XXL and eROSITA samples, respectively.

To predict kSZ profiles from the model, we convert the gas density profile $\rho_{\mathrm{gas}}(r)$ to an electron density profile $n_{\rm e}(r)$ assuming a fully ionised medium with primordial abundances
\begin{equation}\label{eq:ne}
    n_{\rm e}(r)=\frac{(X_{\rm H}+1)}{2}\frac{\rho_{\mathrm{gas}}(r)}{m_{\mathrm{amu}}} \, ,
\end{equation}
where $m_{\mathrm{amu}}$ is the atomic mass unit and $X_{\rm H}=0.76$ is the hydrogen mass fraction.  
The electron density $n_{\rm e}(r)$ is then integrated along the line-of-sight to yield the optical depth
\begin{equation}
    \tau_{\rm gal}(\theta)=2\sigma_\mathrm{T}\int n_{\rm e}(\sqrt{l^2+d_{\rm A}(z)^2\theta^2})dl \,,
\end{equation}
where $d_{\rm A}$ is the angular diameter distance.  
To mirror the observations, we convolve $\tau_{\rm gal}(\theta)$ with a Gaussian beam and employ CAP filtering.
For the DESI samples, the CMB temperature map is convolved with a Gaussian beam of  $1.6'$ FWHM prior to stacking \citep{Ried2025}. 
For the SDSS samples, the 90 and 150 GHz beams are approximately Gaussian with FWHM of $2.1'$ and $1.3'$, respectively \citep{Schaan2021}.
The amplitude of the kSZ effect profile is determined by the standard deviation of the line of sight gas velocities (Equation~\ref{eqn:ksz_stack}). 
We therefore scale the model kSZ profiles by the reported standard deviation of the galaxy peculiar velocities $v^\mathrm{rms}_\mathrm{p}/r_v$ \citep{Schaan2021,Ried2025}; in Appendix~\ref{app:two_halo} we find that $v^\mathrm{rms}_\mathrm{p}/r_v$ is in agreement with the selection of FLAMINGO galaxies that best fits the measured GGL for each kSZ sample.
By multiplying the kSZ profiles by $v^\mathrm{rms}_\mathrm{p}/r_v$, we assume that the standard deviation of the line of sight gas velocities is independent of angular separation from the galaxies $\theta_d$. However, in Appendix~\ref{app:two_halo} we demonstrate that this approximation is only valid for small angular separations. 
Our kSZ fits are therefore limited to $\theta_d<3'$.

\begin{figure*}
\centering
\includegraphics[width=\textwidth]{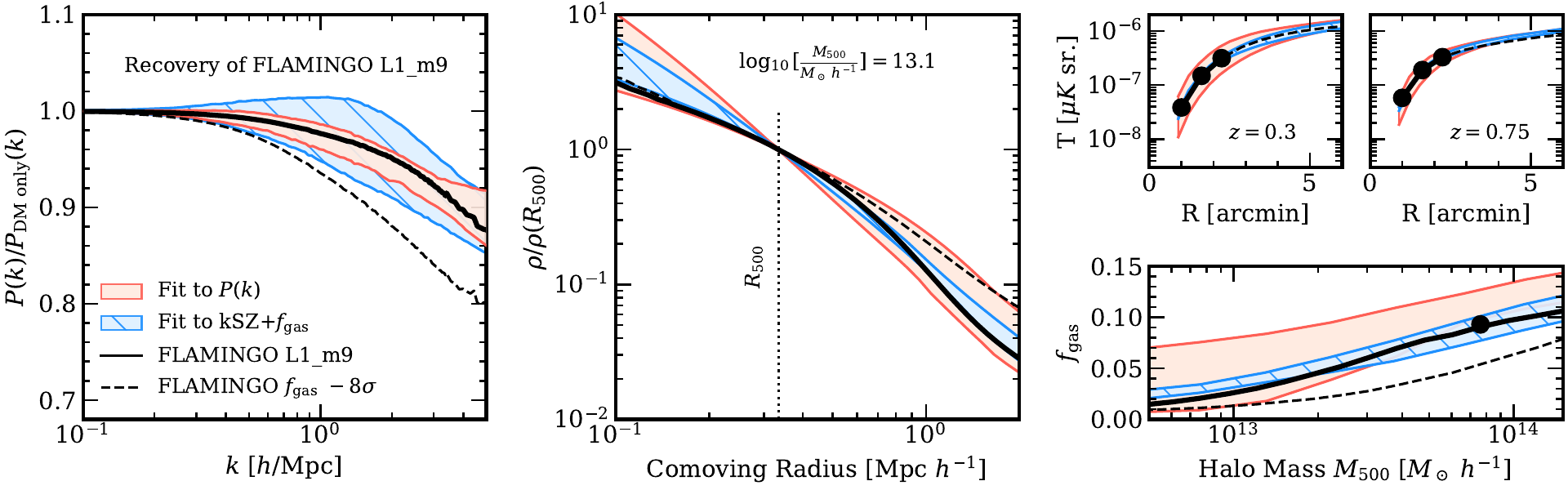}
\includegraphics[width=\textwidth]{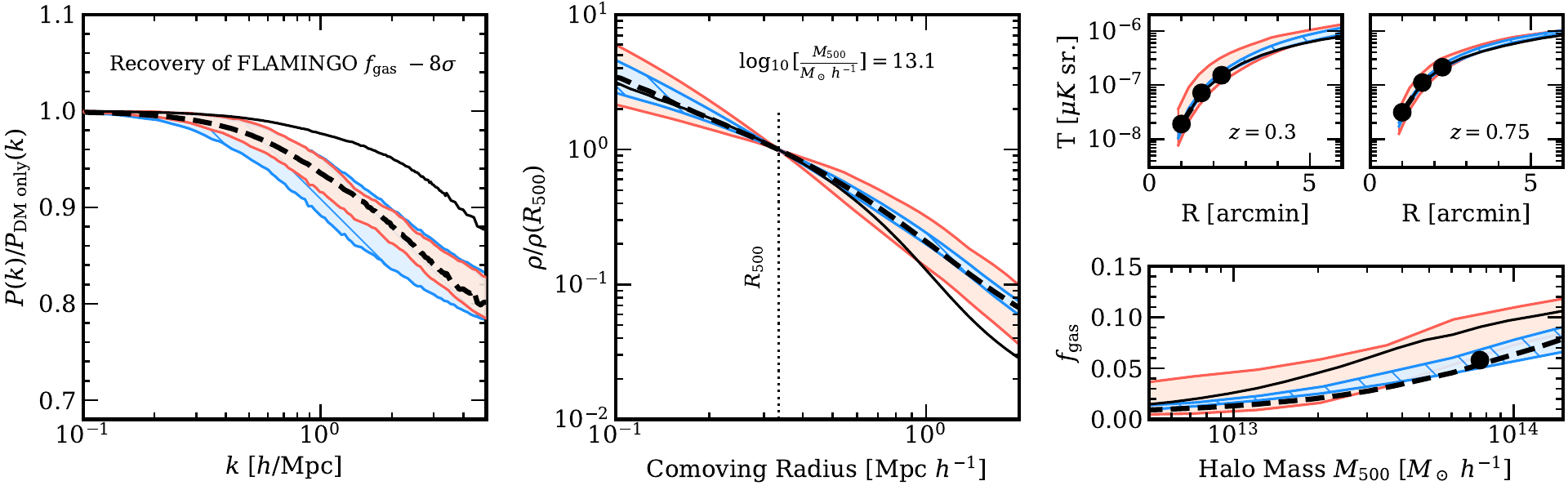}
\caption{Injection recovery test: the baryonification model successfully predicts the suppression of the matter power spectrum, {\pk}, from mock kSZ and gas fraction observations, and vice versa.
The fits to mock kSZ and gas fraction observations are shown in blue, and the fits to {\pk} are shown in pink; the shaded bands represent the $16$ and $84$th percentiles. 
The results for the fiducial (L1\_m9) FLAMINGO simulation (solid black line) are shown in the \textit{top} row, and for the strongest feedback ({\eightsigma}) FLAMINGO simulation (dashed black line) in the \textit{bottom} row.
For each fit, we show {\pk} (\textit{left}), the normalized hot gas radial density profile for a group mass halo (\textit{centre}), the kSZ signal for the same representative halo mass at $z=0.3$ and $0.75$ (\textit{upper right}), and the hot gas mass fraction as a function of halo mass at $z=0$ (\textit{lower right}).
The mock kSZ and gas fraction observations are shown as black circles, with uncertainties of $15\%$ and $5\%$ respectively; the uncertainties are smaller than the markers.
}
\label{fig:injrecovery1}
\end{figure*}
\section{Model validation: Injection--recovery }
\label{sec:inj}

In this work, we infer the suppression of the matter power spectrum relative to a dark matter-only simulation by jointly fitting kSZ and X-ray observations with the baryonification model.
Previously, \cite{Giri2021} demonstrated that the baryonification model successfully predicts the suppression of the matter power spectrum from the mean $f_\mathrm{gas}-M_{500}$ and $f_\mathrm{star}-M_{500}$ relations for the BAHAMAS, OWLS, Illustris, and Horizon simulations. 
The model was also able to predict the gas and stellar mass fractions from the matter power spectrum. 
To further validate the model, we use the FLAMINGO simulations to perform a series of injection--recovery tests designed to mirror our kSZ and X-ray samples.

The FLAMINGO suite consists of $16$ hydrodynamical simulations of varied resolution, box size, subgrid modelling, and cosmology \citep{Schaye2023,Kugel2023}. 
We consider the $1$~Gpc$^3$ intermediate resolution simulations ($m_{\mathrm{gas}} = 1.09\times10^9$ M$_\odot$) with the maximum likelihood cosmological parameters of the DES~Y3 `3×2pt + All Ext.' flat $\Lambda$CDM cosmology \citep{DES3x2}.
The subgrid physics (stellar and AGN feedback) is calibrated to reproduce the observed $z=0$ galaxy stellar mass function and the gas mass fractions of groups and clusters from pre-eROSITA X-ray surveys \citep{Kugel2023};
the calibrated simulations successfully predict the stellar to halo mass relation, the central black hole--stellar mass relation, the cosmic star formation rate history, and cluster scaling relations \citep{Schaye2023,Braspenning2024}.
The suite includes variants with stronger (weaker) baryon feedback that were calibrated to gas fraction measurements shifted down (up) by $ N\sigma$, where $\sigma$ is the observational uncertainty on the mean gas fraction relation and $N \in [+2,-2,-4,-8]$. 
For our tests, we consider the fiducial feedback (L1\_m9) and strongest feedback ({\eightsigma}) simulations.
Both simulations use thermal AGN feedback following \cite{Booth2009}; the {\eightsigma} simulation realizes stronger feedback by more powerful but less frequent AGN outbursts.

\subsection{Predicting the gas observables from the power suppression}

We first assess whether the baryonification framework can predict the properties of halos from the suppression of the matter power spectrum,  $P(k) / P_\mathrm{DM~Only}(k)$.
For the fiducial and strongest feedback FLAMINGO simulations, we fit $P(k) / P_\mathrm{DM~Only}(k)$ at $z=0$ and $0.1 < k < 5~h~\mathrm{Mpc}^{-1}$.
We adopt uniform priors for all seven free parameters and assume $5\%$ uncertainty in $P(k) / P_\mathrm{DM~Only}(k)$.
The parameter posteriors are shown in Appendix~\ref{app:injrec}.

The model successfully fits the shape of the power suppression for both simulations (left column of Figure~\ref{fig:injrecovery1}).
From only $P(k) / P_\mathrm{DM~Only}(k)$, the fits accurately predict the baryon content of groups and clusters. 
In Figure~\ref{fig:injrecovery1}, we compare the gas mass fractions of groups and clusters with the model predictions, finding strong agreement;
the model also reproduces the stellar fractions, shown in Appendix~\ref{app:injrec} for brevity. 
In addition to integrated properties (e.g., $f_\mathrm{gas}$), the model also accurately predicts the radial extent of the gas.
For a representative group mass halo ($\log_{10} M_{500} / \mathrm{M}_\odot~h^{-1} = 13.1$), we compare the gas density profile and the kSZ effect from the simulations with the model predictions (Figure~\ref{fig:injrecovery1}); the model and simulation are consistent within $1\sigma$.

\begin{figure*}
\centering
\includegraphics[width=\textwidth]{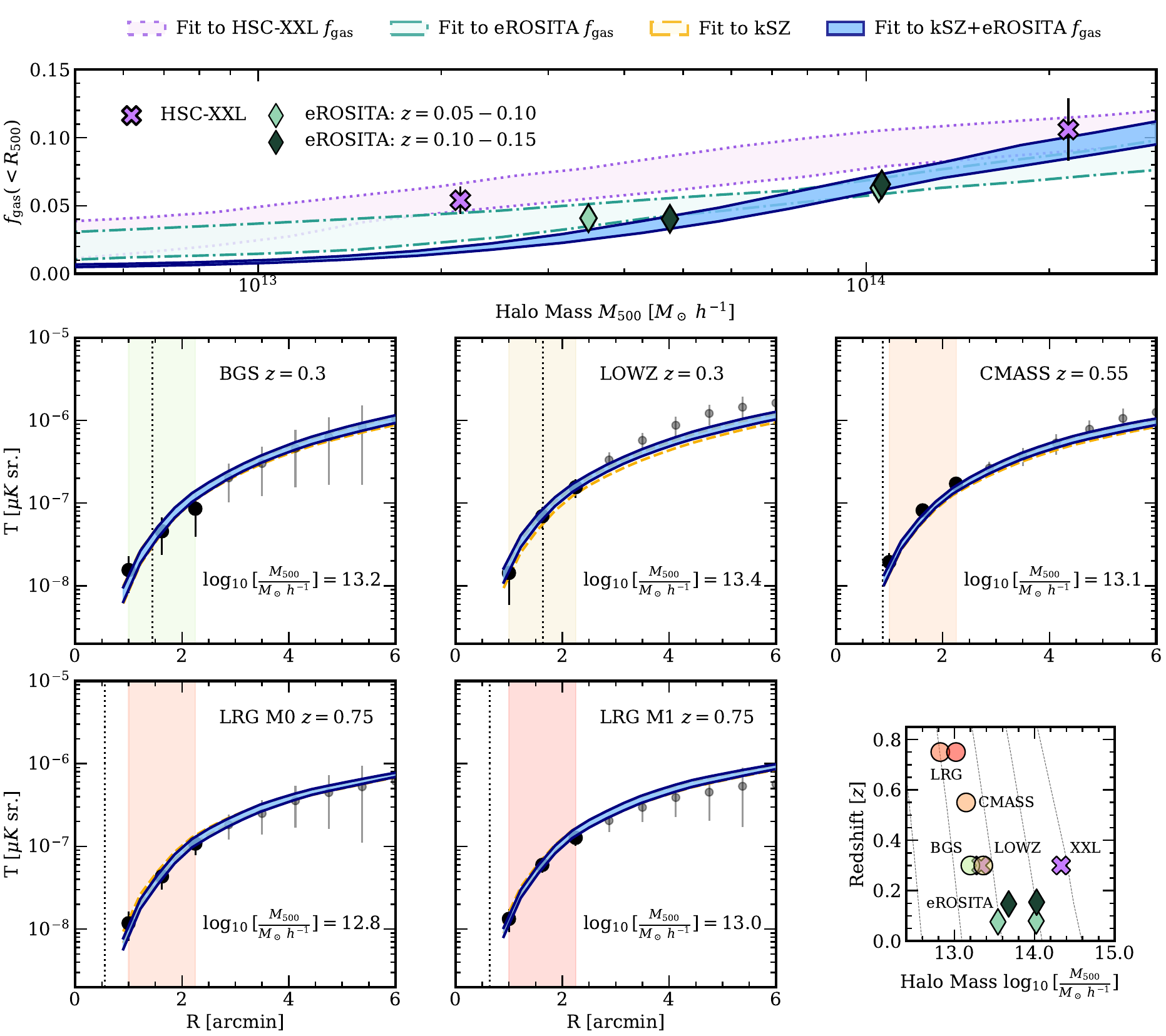}
    \caption{The joint fit to the kSZ and eROSITA X-ray measurements (blue), alongside fits to the kSZ (yellow), eROSITA (green), and HSC--XXL (purple) data separately.
    The top row compares the X-ray gas mass fraction measurements with the X-ray-only fits, and the lower two rows compare the kSZ effect profiles with the kSZ-only fit;
    the joint fit is presented in all panels.
    The shaded bands represent the $16$ and $84$th percentiles of the model fits.
    For the kSZ measurements, we restrict our fits to the innermost data points ($<3'$), as indicated by the vertical shaded bars; at these scales the data is most constraining and least impacted by uncertain modelling (Appendix~\ref{app:two_halo}). 
    For each kSZ observation, we demarcate $R_{500}$ as a vertical dotted line.
    The lower right panel presents the mean halo mass and redshift of each observable. 
    The dotted lines show the mean growth histories of halos, binned by their redshift zero mass.
    }
    \label{fig:goodness_of_fit_mega}
\end{figure*}

\begin{figure*}
\centering
\includegraphics[width=\textwidth]{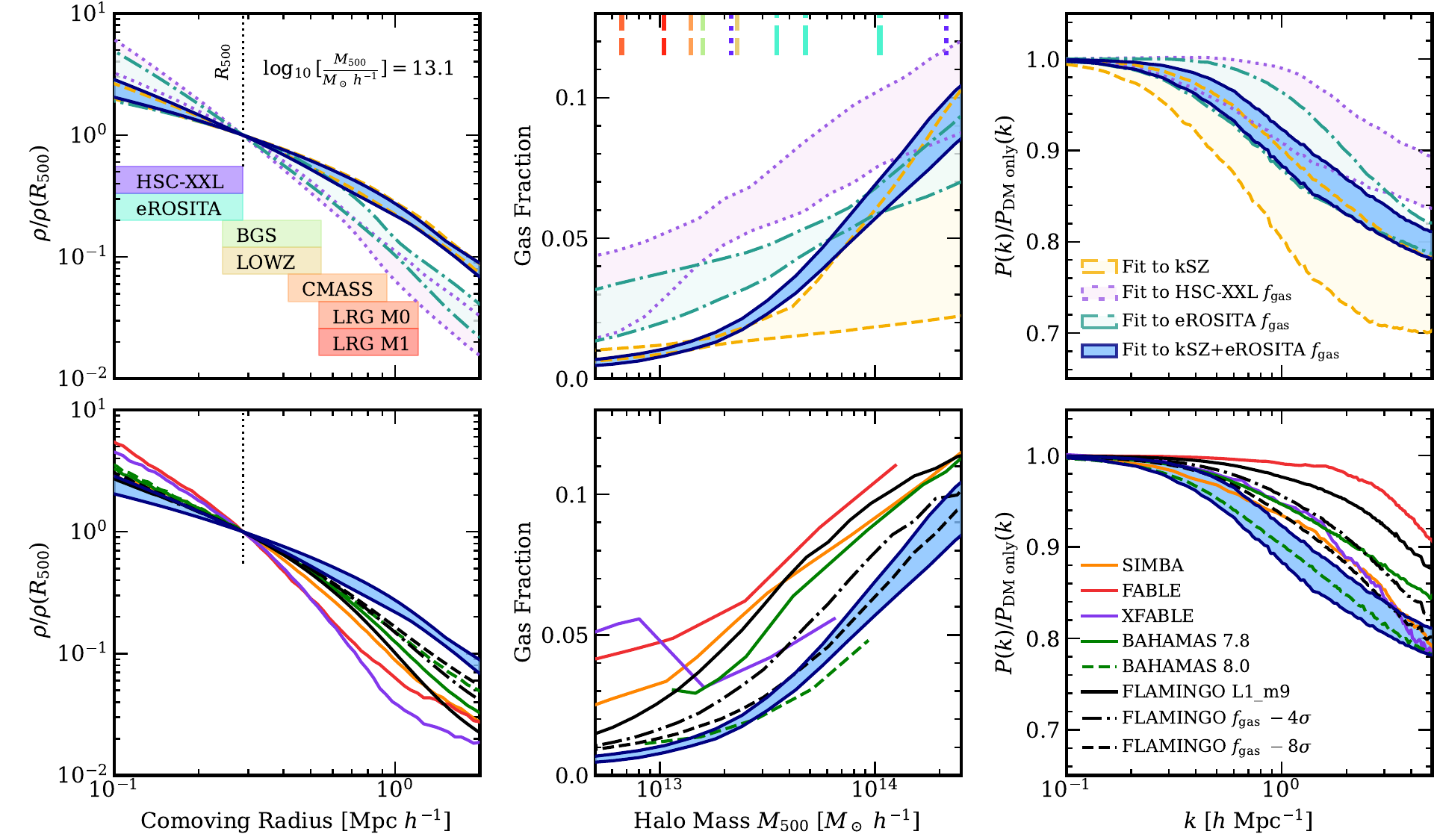}
    \caption{
    Constraints from  
    SDSS/DESI$+$ACT kSZ effect profiles and X-ray gas fractions on the radial density profile of a group mass halo (\textit{left}), the hot gas mass fraction as a function of halo mass (\textit{centre}), and the suppression of the total matter power spectrum (\textit{right}). 
    \textit{Top}: the results of the joint kSZ and eROSITA X-ray fit (blue), alongside fits to the kSZ (gold), eROSITA (green), and HSC--XXL (purple) data separately.
    We note that the kSZ-only fit is prior dominated at cluster masses, while the X-ray fits are prior dominated at group masses.
    The shaded bands correspond to the $16$ and $84$th percentiles.
    In the leftmost panel, the range of comoving radii probed by each observation is indicated by the horizontal shaded bars.
    In the centre panel, the observations' mean halo masses are shown as vertical lines (dashed for kSZ, dot-dashed for eROSITA, and dotted for HSC--XXL).
    \textit{Bottom}: 
    the joint kSZ and eROSITA fit alongside a selection of recent hydrodynamical simulations: BAHAMAS \citep{McCarthy2017}, SIMBA \citep{Dave2019}, FLAMINGO \citep{Schaye2023}, FABLE/XFABLE \citep{Henden2018,Bigwood2025XFABLE}.
    }
    \label{fig:fgas_and_pk}
\end{figure*}

\subsection{Predicting the power suppression from the gas observables}

We infer the power suppression of the FLAMINGO simulations by fitting mock kSZ effect profiles and hot gas fractions with the baryonification framework.
The simulated kSZ effect profiles are calculated from the FLAMINGO lightcones following \cite{McCarthy2025}.
To reflect the SDSS/DESI+ACT kSZ measurements  \citep[][]{Schaan2021,Ried2025}, 
we stack the kSZ effect profile for group mass halos $13.1<\log_{10} M_{500}/ \mathrm{M}_\odot~h^{-1}<13.2$ at $z=0.3$ and $0.75$, using a Gaussian beam of FWHM$=1.6'$ and CAP filtering.
For the gas mass fraction observables, we calculate $f_\mathrm{gas} = M_\mathrm{gas} / M_{500}$ from the stacked density profile of $z=0.3$ clusters: $13.95<\log_{10} M_{500}/ \mathrm{M}_\odot~h^{-1}<14.05$.
We assume uncertainties of $15$ and $5\%$ on the simulated kSZ and gas mass measurements, respectively; these uncertainties are approximately twice the signal-to-noise ratio of the real data.

For each simulation, we fit the mock measurements with the baryonification model.
Mirroring our analysis of the real data, we do not include stellar fractions in our fiducial simulation test.
We demonstrate successful recovery of the power suppression from joint $f_\mathrm{gas}+f_\mathrm{star}+$kSZ fits in Appendix~\ref{app:injrec}.
The joint kSZ and gas fraction fits are presented in Figure~\ref{fig:injrecovery1}.
For both simulations, the fit successfully describes the mock observables and accurately predicts the power suppression (within $1 \sigma$).
The posteriors of the fits are included in Appendix~\ref{app:injrec}.

\section{Results: Joint kSZ and X-ray fits}
\label{sec:results}

We constrain the matter power spectrum from kSZ and X-ray observations of the gas distribution (Section~\ref{sec:data}): five stacked SDSS/DESI+ACT kSZ effect profiles of group mass halos at $0.2<z<0.8$  \citep{Schaan2021,Ried2025} and X-ray gas mass measurements of clusters from eROSITA \citep[$z\approx 0.1$,][]{Bulbul2024} and HSC--XXL \citep[$z\approx 0.3$,][]{Akino2022}.
The mean halo mass of each sample is characterized with weak lensing: \cite{Siegel2025flamingo} for the kSZ and eROSITA samples, and \cite{Umetsu2020} for HSC--XXL.

The data is fit with the baryonification model, using wide top hat priors for the five gas parameters ($M_{\rm c}, \mu, \theta_{\rm ej}, \delta, \gamma$). 
The posterior probability distribution is sampled with \texttt{emcee} \citep{ForemanMackey2013}.
We account for several  sources of uncertainty in our analysis:
\begin{enumerate}
\item Mean halo mass: for the kSZ and eROSITA samples, we adopt Gaussian priors on the halo mass, with the centres and standard deviations set by the GGL measurements;
for the HSC--XXL data, halo mass uncertainties were propagated into the gas mass fractions through the Bayesian modelling of \cite{Akino2022}.
\item X-ray gas mass: the statistical uncertainty on the mean gas mass of each eROSITA bin is $<1\%$, however, we adopt a conservative $10\%$ uncertainty to reflect potential systematics \citep[e.g., the $15\%$ flux offset between eROSITA and \textit{Chandra} on common SPT sources, which approximately corresponds to $7\%$ in gas mass,][]{Bulbul2024}.
\item kSZ effect profiles: for the kSZ effect profiles, we consider the measurements' full covariance matrices. 
The amplitude of the kSZ effect is proportional to the RMS of the line-of-sight peculiar velocities, for which \cite{Hadzhiyska2024} report $<5\%$ uncertainty;
we assume a conservative uncertainty of $10\%$.
\end{enumerate}

\subsection{Consistency of the kSZ and X-ray data} 
\label{sec:consistency}

First, we consider the three datasets independently, to assess their consistency and the feasibility of a joint fit: 
i) SDSS/DESI+ACT kSZ effect profiles, ii) eROSITA gas mass fractions, and iii) HSC--XXL gas mass fractions.
The fits are presented alongside the data in Figure~\ref{fig:goodness_of_fit_mega}.
The posteriors of each fit are reported in Table~\ref{tab:priors} and shown in Figure~\ref{fig:corner_data}.

The kSZ effect profiles and the eROSITA gas mass fractions favour similar feedback strengths, consistent with previous findings \citep{Kovac25,Siegel2025flamingo}.
The posteriors of the kSZ and eROSITA fits are consistent at the $1\sigma$ level for all parameters, allowing their joint analysis.
Because the kSZ-only and eROSITA-only fits probe unique and relatively narrow ranges of halo mass, their predictions for the gas mass fraction relation should be interpreted with caution (Figure~\ref{fig:fgas_and_pk});
the kSZ-only fit is prior dominated at cluster masses, while the eROSITA-only fit is prior dominated at group masses. 
The joint fit to the kSZ and eROSITA data is presented in Figure~\ref{fig:goodness_of_fit_mega} and Table~\ref{tab:priors}.
Across group and cluster masses ($13<\log_{10} \frac{M_{500}}{ \mathrm{M}_\odot~h^{-1}} <14$) and at $0<z<1$, the model describes the data well: reduced $\chi^2=1.2$. 
\footnote{
We also calculate the reduced $\chi^2$ using the effective number of parameters: $p_D = -2 \langle \log p(\mathrm{data} \mid \theta) \rangle  +2 \log p(\mathrm{data} \mid \langle \theta \rangle)$, where $p(\mathrm{data} \mid \theta)$ is the posterior probability of the parameters $\theta$ and the averages are over the MCMC posterior samples \citep{spiegelhalter2002bayesian}.
For the joint kSZ and eROSITA fit, we find $p_D=10$ and  $\chi^2/p_D=0.9$.
}

The HSC--XXL X-ray measurements imply relatively weaker gas expulsion than the kSZ and eROSITA measurements; 
in particular, the kSZ and HSC--XXL fits differ at the $2\sigma$ level for $M_\mathrm{c}$---the halo mass below which the gas density profile becomes shallower (Equation~\ref{eq:beta}).
The baryonification model gives a poor joint fit to these data: reduced $\chi^2=2.5$.\footnote{
For the joint kSZ and HSC--XXL fit, $p_{\rm D}=10$ and $\chi^2/p_{\rm D}=2.1$.
} The joint model is presented against the data in Figure~\ref{fig:goodness_of_fit_mega_akino}.
Motivated by the tension between the CMASS kSZ measurement and the pre-eROSITA X-ray samples \citep{bigwood2024}, previous studies proposed that the apparent discrepancy could result from the different radial scales, mean halo masses, and redshifts of the samples \citep{bigwood2024,LucieSmith2025}.
However, the SDSS LOWZ and DESI BGS kSZ measurements favour more gas expulsion than the lower mass bin of HSC--XXL, despite the samples having similar mean halo masses and redshifts (Figure~\ref{fig:goodness_of_fit_mega});
at this mass ($\log_{10}M_{500}/\mathrm{M}_\odot~h^{-1}\approx13.3$) and redshift ($z\approx0.3$), the kSZ measurements probe the same radii as the X-ray measurements ($<R_{500}$).
The disagreement between the kSZ measurements and the HSC--XXL scaling relation could stem from unaccounted for systematics in the X-ray or kSZ data (see Section~\ref{sec:xray_disc}).
Simulation studies of the X-ray and kSZ measurements are warranted to investigate observational and modelling systematics. 

Encouraged by the consistency of the data in analytic modelling and simulation comparisons, we primarily consider the joint kSZ and eROSITA model.
However, we also investigate the HSC--XXL-only fit to reflect the uncertain state of the X-ray measurements.

\subsection{Gas properties of groups and clusters} 

The kSZ and X-ray measurements probe a wide range of redshifts, halo masses, and radii: the kSZ measurements probe the gas distribution out to several $R_{500}$ of group mass halos at $0.3<z<1.0$, and the X-ray measurements are most informative for the inner regions ($<R_{500}$) of low redshift clusters ($z<0.3$).
The range of radii probed by each measurement is highlighted in Figure~\ref{fig:fgas_and_pk}. 
Together, the measurements probe all radii between $0$ and $1$~comoving Mpc.

In Figure~\ref{fig:fgas_and_pk}, we present the model constraints on the gas properties of groups and clusters: the radial density profile of a representative group mass halo at $z=0$ and the hot gas mass fraction scaling relation. 
We show the results from fitting the kSZ, eROSITA, and HSC--XXL data separately, as well as the joint kSZ and eROSITA X-ray fit.
As discussed in Section~\ref{sec:consistency}, the feedback constraints from the kSZ and eROSITA data are broadly consistent, but the HSC--XXL X-ray data favours weaker gas expulsion. 
It is difficult to quantitatively compare the gas property constraints from the separate kSZ and X-ray fits, because the kSZ-only fit is prior dominated at cluster masses, while the X-ray fits are prior dominated at group masses.  The consistency of the independent fits is best ascertained from their posteriors and the quality of a joint fit.

The joint kSZ and eROSITA fit is highly constraining and yields precise predictions for the gas properties of groups and clusters, highlighting the benefit of constraining the gas distribution across a wide range of halo mass and redshift. In Section~\ref{sec:implications_for_feedback}, we compare the model fits with recent hydrodynamical simulations.

\subsection{Suppression of the matter power spectrum}

With the \texttt{BCemu} emulator \citep{Giri2021}, we map between the density profile parameters ($M_{\rm c}, \mu, \theta_{\rm ej}, \delta, \gamma, \eta, \eta_\delta$) and the suppression of the matter power spectrum.
The posteriors on {\pk} from the independent kSZ, eROSITA, and HSC--XXL fits are presented in Figure~\ref{fig:fgas_and_pk}, as well as our joint kSZ and eROSITA model.

Each gas observable favours suppression of the matter power spectrum relative to a dark matter-only simulation at $k>1~h~\mathrm{Mpc}^{-1}$.
The {\pk} constraints from fitting the kSZ and eROSITA X-ray data separately are statistically consistent with each other, and both are consistent with the joint fit at the $1\sigma$ level (Figure~\ref{fig:fgas_and_pk}).
The posterior on {\pk} is significantly narrowed by jointly fitting the kSZ and X-ray measurements.
The model favours strong suppression of the matter power spectrum relative to a dark matter-only simulation: $10\pm 2\%$ at $k=1~h~\mathrm{Mpc}^{-1}$. Compared to the kSZ and eROSITA data, the HSC--XXL scaling relation implies weaker 
suppression.
We compare our {\pk} constraint with previous measurements and simulation predictions in Sections~\ref{sec:other_feedback_constraints} and~\ref{sec:implications_for_feedback}, respectively.

\begin{figure*}
\centering
\includegraphics[width=\textwidth]{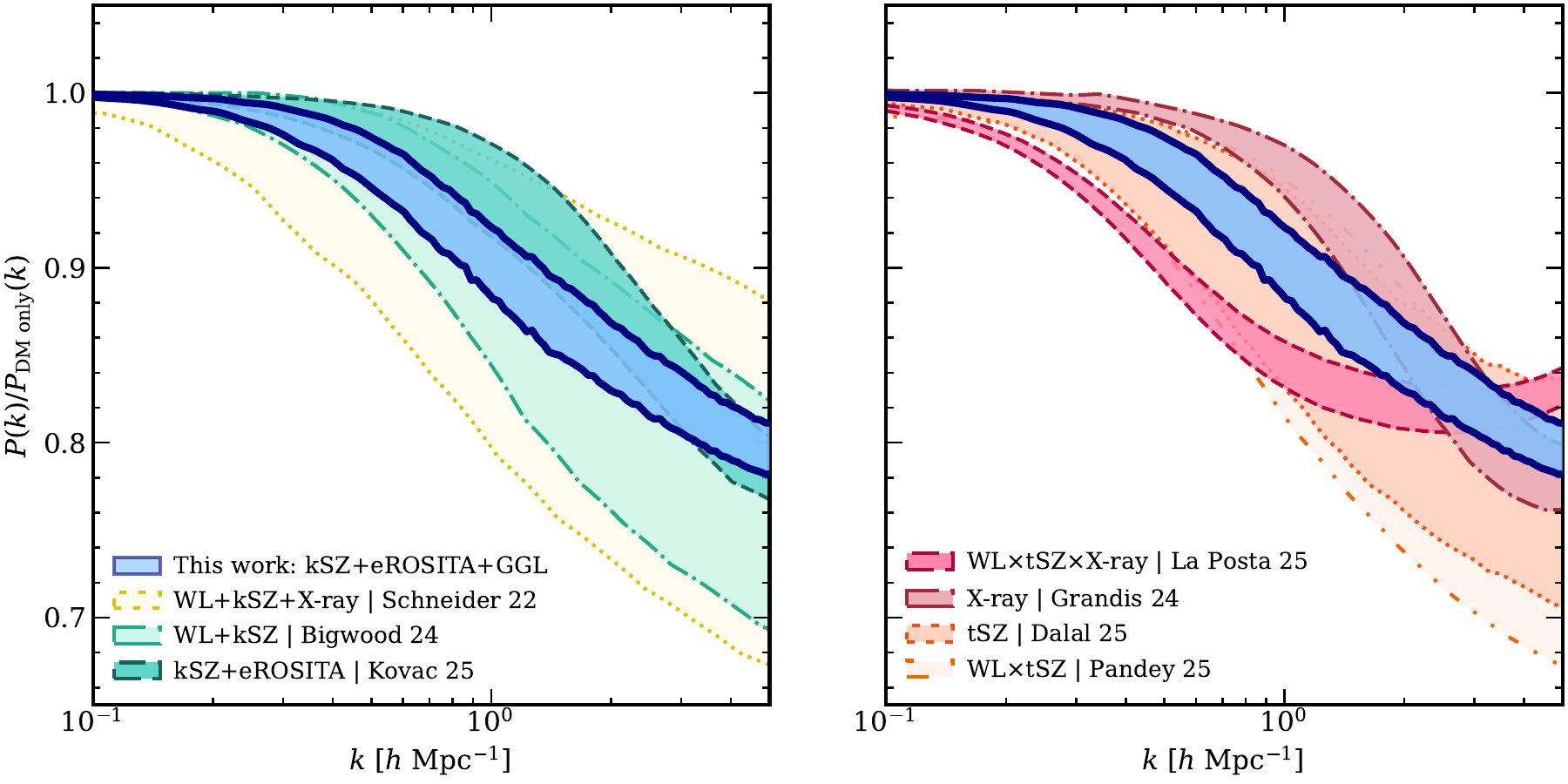}
    \caption{
    The constraints on the suppression of the matter power spectrum {\pk} from our joint fit to kSZ effect profiles and eROSITA X-ray gas mass fractions (blue), compared to constraints from the literature.
    \textit{Left}: constraints derived (in part) from kSZ effect measurements \citep{Schneider2022,bigwood2024,Kovac25}.
    \textit{Right}: constraints from pre-eROSITA X-ray and/or tSZ measurements \citep{Grandis2024,LaPosta2025,Dalal2025,Pandey2025}. }
    \label{fig:Pk_Map}
\end{figure*}

\section{Discussion}
\label{sec:disc}

\subsection{Uncertainties in the X-ray data}
\label{sec:xray_disc}

By probing the amount of gas depletion within $R_{500}$ of groups and clusters,  X-ray gas mass measurements are powerful probes of baryon feedback.
In this work, we consider X-ray gas mass measurements from the eROSITA all-sky survey \citep{Bulbul2024} and the HSC--XXL sample \citep{Pierre2016,Eckert2016,Umetsu2020,Akino2022}.
The eROSITA observations suggest that groups and clusters are on average more gas depleted than indicated by the HSC--XXL scaling relation \citep{Akino2022}.
The average eROSITA gas mass fractions are also lower than those found by \cite{Kugel2023} from a collection of pre-eROSITA samples (without selection effect corrections):
\cite{Vikhlinin2006,Maughan2008, Rasmussen2009,
Sun2009,
Pratt2010,
Lin2012,
Lagana2013,
Sanderson2013,
Gonzalez2013,
Lovisari2015,
Hoekstra2015,
Pearson2017,
Mulroy2019,
Lovisari2020}.
This shift to lower mean gas mass fractions in the eROSITA data relative to pre-eROSITA samples was noted by several previous studies \citep{Popesso2024,Dev2024,Kovac25,Siegel2025flamingo}.

Selection effects are a potential explanation for the apparent discrepancy between the eROSITA measurements and the earlier samples \citep[e.g.,][]{Seppi2022,Popesso2024a,Dev2024,Marini2024,Clerc2024,Seppi2025}.
X-ray luminosity is proportional to density and gas mass (along with a temperature and metallicity dependence), causing X-ray selected samples to be biased to concentrated gas rich halos \citep{Andreon2017,Seppi2022,Marini2024,Clerc2024,Seppi2025}.  
The lower mean gas mass fractions found by eROSITA could therefore reflect its greater completeness compared to previous X-ray selected samples \citep[e.g., ROSAT,][]{Truemper1982,Voges1999}. 
However, the HSC--XXL scaling relation analysis accounts for selection effects but still favours higher average gas mass fractions than reported by eROSITA.
\cite{Akino2022} model the X-ray selection effect by truncating the parent distribution with a minimum X-ray luminosity threshold.
The disagreement between the HSC--XXL scaling relation and the recent eROSITA measurements suggests that selection effects are not the only factor behind the discrepancy and/or the minimum luminosity threshold correction is insufficient \citep[e.g.,][]{Andreon2017}. 

The eROSITA sample is also shaped by selection effects, with a significant drop in completeness for group mass halos and below \citep[e.g.,][]{Marini2024}.
To bypass X-ray selection, \cite{Popesso2024} used eROSITA images to measure the gas mass fractions for stacks of optically selected groups and clusters \citep{Robotham2011,Driver2022}. 
These measurements mitigate the impact of X-ray selection but are subject to optical group finder contamination \citep[$>30\%$,][]{Seppi2025} and halo mass uncertainties.
\cite{Pearson2017} similarly studied a sample of $10$ groups from the GAMA friends-of-friends catalogue with \textit{Chandra}.
The mean gas mass fractions of our eROSITA cluster bins, which are limited to $z<0.2$ to increase completeness, are consistent with these optically selected samples.
Stronger feedback than previously assumed is also favoured from a comparison of the entropy of eROSITA detected groups with cosmological hydrodynamic simulations (MillenniumTNG, Magneticum, OWLS), including the effects of the eROSITA selection function \citep{Bahar2024}.
Studying the impact of the eROSITA selection function on the sample of gas mass fraction measurements is critical to understanding the extent of baryon feedback (Ding et al. in prep).

The apparent discrepancy in mean gas mass fractions could stem from biased halo mass estimation.
Previous studies often assumed hydrostatic equilibrium to estimate total masses, and such estimates can be biased by approximately $30\%$ due to non-thermal pressure support and/or deviations from hydrostatic equilibrium \citep{Hoekstra2015,Eckert2016,Kugel2023,MunozEcheverria2024,Braspenning2025}.
However, both the HSC--XXL scaling relation and the bins of eROSITA clusters that we consider include weak lensing halo mass constraints, from \cite{Umetsu2020} and \cite{Siegel2025flamingo}, respectively.

Systematics in X-ray modelling also warrant further investigation.
For a sample of SPT clusters, eROSITA derived luminosities are approximately $15\%$ lower than previously inferred with \textit{Chandra} \citep{Bulbul2024}.
Because X-ray luminosity is proportional to the gas density squared (at fixed temperature and metallicity), this flux offset corresponds to a  $7\%$ skew in the gas mass. 
We include this source of uncertainty in our modelling (Section~\ref{sec:results}).

\subsection{Landscape of feedback constraints}
\label{sec:other_feedback_constraints}

We present a selection of {\pk} measurements, alongside our joint kSZ and eROSITA result in Figure~\ref{fig:Pk_Map} \citep{Schneider2022,bigwood2024,Grandis2024,LaPosta2025,Pandey2025,Kovac25,Dalal2025}.
For brevity, we limit our comparison to studies that considered gas observables (e.g., X-ray and kSZ); hydrodynamical simulation predictions are shown in Figure~\ref{fig:fgas_and_pk} and discussed in Section~\ref{sec:implications_for_feedback}.
The joint kSZ and eROSITA X-ray {\pk} constraint is consistent with the collection of previous studies at approximately the $1 \sigma$ level \citep[with the exception of][discussed below]{LaPosta2025}.
The {\pk} constraints derived solely from the pre-eROSITA gas mass fractions \citep[our HSC--XXL fit and][]{Grandis2024} favour slightly weaker power suppression than the other literature constraints and our joint kSZ and eROSITA model; however, the pre-eROSITA constraints are still broadly consistent with the other {\pk} posteriors.

The consistency of the recovered {\pk} between these studies is particularly encouraging because each observational probe is sensitive to different redshifts and halo masses. 
Several studies \citep{Schneider2022,bigwood2024,Kovac25} analysed the CMASS kSZ effect profile \citep[$z\approx0.55$ and $\log_{10}M_{500}/\mathrm{M}_\odot \approx 13$,][]{Schaan2021}, along with a selection of pre-eROSITA X-ray gas fraction measurements between $13<\log_{10}M_{500}/\mathrm{M}_\odot<15$ at $z\lesssim0.3$, including \cite{Vikhlinin2009, Gonzalez2013, Sanderson2013, Lovisari2015, Kravtsov2018, Akino2022, Popesso2024};
the pre-eROSITA cluster hot gas mass mass fractions and radial density profiles analysed by \cite{Grandis2024} cover a similar mass and redshift range.
The cross correlation of the tSZ effect with cosmic shear is also sensitive to $M_{500}\approx10^{14} \mathrm{M}_\odot$ halos but over a wider redshift range  \citep[$z<1$,][]{Hojjati2015,Pandey2022,Pandey2025}.
The sample of tSZ selected clusters in \cite{Dalal2025} have a mean redshift of $z = 0.55$ and a mean halo mass of $ M_{500}  = 3 \times 10^{14}~\mathrm{M}_\odot$.
In this work, we jointly analyse a selection of GGL calibrated kSZ and X-ray samples between group and cluster masses at $0<z<1$; 
the stacked measurements of the kSZ effect from DESI$+$ACT are vital for probing $z > 0.5$ \citep{Ried2025}.
By spanning a wide range of halo mass and redshift, joint modelling of kSZ and X-ray measurements significantly improves the constraining power of the model fit (Figures~\ref{fig:fgas_and_pk} and \ref{fig:corner_data}).
Although our model considers the widest range of halo mass and redshift thus far, the inclusion of additional observables, particularly the unique mass and redshift space probed by the tSZ samples, is critical to building a complete picture of baryon feedback.

A quantitative assessment of the consistency of our results with those in Figure~\ref{fig:Pk_Map} is limited, because different models have been used to constrain the suppression of the matter power spectrum. 
In this work, we adopt the \texttt{BCemu} emulator \citep{Giri2021}; several previous studies also considered \texttt{BCemu} \citep{Schneider2022,bigwood2024,Grandis2024} or similar \texttt{BACCO} emulators \citep{Arico2023,Grandis2024}. 
\cite{Kovac25} utilized the next generation baryonification model recently introduced by \cite{Schneider2025}; 
the improved baryonification model includes updated formulations of the density profiles and treatment of the dark matter, gas, and stellar components as independent fields (an emulator is not yet publicly available).
Other results use the halo model approach \citep{Debackere2020,LaPosta2025,Pandey2025,Dalal2025}. 
The power suppression constraints from the baryonification and halo model frameworks are largely consistent. 
The {\pk}  posterior from \cite{LaPosta2025} marginally deviates from the other constraints, potentially reflecting the relatively lower flexibility of their assumed density profiles (e.g., fixed ejection radius of unbound gas).
As observational probes of baryon feedback become more precise, modelling systematics are increasingly significant. 
Future work comparing methods of constraining the suppression of the matter power spectrum and addressing modelling challenges (e.g., the effect of satellites and miscentering on the kSZ effect profile, Appendix~\ref{app:two_halo}) is warranted.

\subsection{Implications for feedback models}
\label{sec:implications_for_feedback}

Cosmological hydrodynamical simulations reproduce a variety of galaxy and cluster observables, including the stellar-to-halo mass relation and cluster scaling relations \citep[e.g.,][]{McCarthy2017, Dave2019, Schaye2023, Pakmor2023, Schaye2025,Dolag2025},
however, their predictions for the extent of gas expulsion from groups and clusters and the suppression of the matter power spectrum differ substantially.
In Figure~\ref{fig:fgas_and_pk}, we present the gas mass fractions of groups and clusters and the suppression of the matter power spectrum from a selection of recent simulations: BAHAMAS \citep{McCarthy2017}, SIMBA \citep{Dave2019}, FLAMINGO \citep{Schaye2023}, FABLE \citep{Henden2018}, and XFABLE \citep{Bigwood2025XFABLE}.
At $k=1~h~\mathrm{Mpc}^{-1}$, the predictions for the suppression of the matter power spectrum differ by $>10\%$.

In modern hydrodynamical simulations, the suppression of the matter power spectrum is primarily regulated by the AGN subgrid model.
SIMBA treats black hole accretion with a two-phase model: torque-limited accretion for cold gas and Bondi accretion for hot gas near the black hole \citep{HopkinsQuataert2011, Angles_Alcazar2017a}; 
AGN feedback is implemented with both kinetic and radiative modes. 
FABLE also employs two-mode AGN feedback \citep[following Illustris,][]{Sijacki2015}:
the quasar-mode (high accretion rates), where the feedback energy is thermally coupled to the surrounding gas, and the radio mode (low accretion rates), where the feedback energy is injected through hot bubbles. 
XFABLE achieves stronger matter power suppression than FABLE, while remaining in agreement with observed galaxy and cluster properties, by modifying the AGN radio mode sub-grid model \citep{Bigwood2025XFABLE}:
the hot bubbles are injected at $100$~kpc$/h$ (compared to $\sim30$~kpc$/h$ in FABLE),  
the accretion threshold for the radio mode is raised by $10\times$, only AGN in massive halos ($M_{500} \gtrsim 10^{13}~\mathrm{M}_{\odot}$) enter the radio mode, and the energy content of the radio bubble is limited to $20\times$ the total energy in the ICM. 
The feedback models of FABLE and XFABLE were calibrated to match pre-eROSITA gas mass fractions;
SIMBA was not calibrated to group and cluster scale properties.
The fiducial FLAMINGO and BAHAMAS simulations implement AGN feedback through the radiative heating mechanism of \cite{Booth2009}.
To avoid numerical overcooling \citep{Dalla2012}, the feedback energy is stored internally until it can heat at least one gas particle by $\Delta T_\mathrm{AGN}$; 
increasing $\Delta T_\mathrm{AGN}$ results in more powerful but less frequent AGN outbursts.  
The BAHAMAS suite includes a strong feedback variant with $\log_{10}\Delta T_\mathrm{AGN}=8.0$, compared to the fiducial BAHAMAS choice of  $\log_{10}\Delta T_\mathrm{AGN}=7.8$. 
The FLAMINGO suite also includes variants with stronger (weaker) baryon feedback by calibrating to pre-eROSITA gas fraction measurements shifted down (up) by $ N\sigma$, where $\sigma$ is the observational uncertainty on the mean gas fraction relation and $N \in [+2,-2,-4,-8]$ \citep{Kugel2023}.
For brevity, we only consider the fiducial intermediate resolution FLAMINGO simulation (L1\_m9) and the strongest feedback variant ({\eightsigma}).
The FLAMINGO suite also includes variants with jet mode AGN feedback \citep{Husko2022}, however, these variants were only produced for a limited range of feedback strengths ($N \in [0,-4]$).

Joint modelling of kSZ and eROSITA X-ray data requires significantly more gas depletion within $R_{500}$ than any of the fiducial feedback strength simulations predict.
Only the simulation variants with particularly strong feedback, BAHAMAS 8.0 and FLAMINGO {\eightsigma}, are consistent with our model.
The comparison of hydrodynamic simulations to recent kSZ and eROSITA observations also favours strong feedback simulation variants \citep{Hadzhiyska2024photoz, Ried2025,McCarthy2025,Siegel2025flamingo,Bigwood2025allthesims}.

The state of the ICM (e.g., temperature and pressure) is also sensitive to the AGN feedback.
\cite{Braspenning2024} found that the fiducial FLAMINGO simulation and the more moderate feedback variants (e.g., $f_\mathrm{gas}~-4\sigma$) are consistent with the observed cluster scaling relations and the radial profiles of massive clusters from a collection of the pre-eROSITA data.
However, the strongest feedback variant ({\eightsigma}) deviates from the cluster scaling relations, despite providing the best match to the kSZ measurements of all FLAMINGO variants. 
Exploring feedback models that can successfully reproduce all these observations simultaneously, including alternate formulations of AGN feedback \citep{Bigwood2025XFABLE} or other mechanisms of gas expulsion \citep[e.g., cosmic ray transport,][]{Quataert2025}, is  warranted.

\subsection{Data-driven priors for weak lensing cosmology}

\begin{figure}
\centering
\includegraphics[width=0.85\columnwidth]{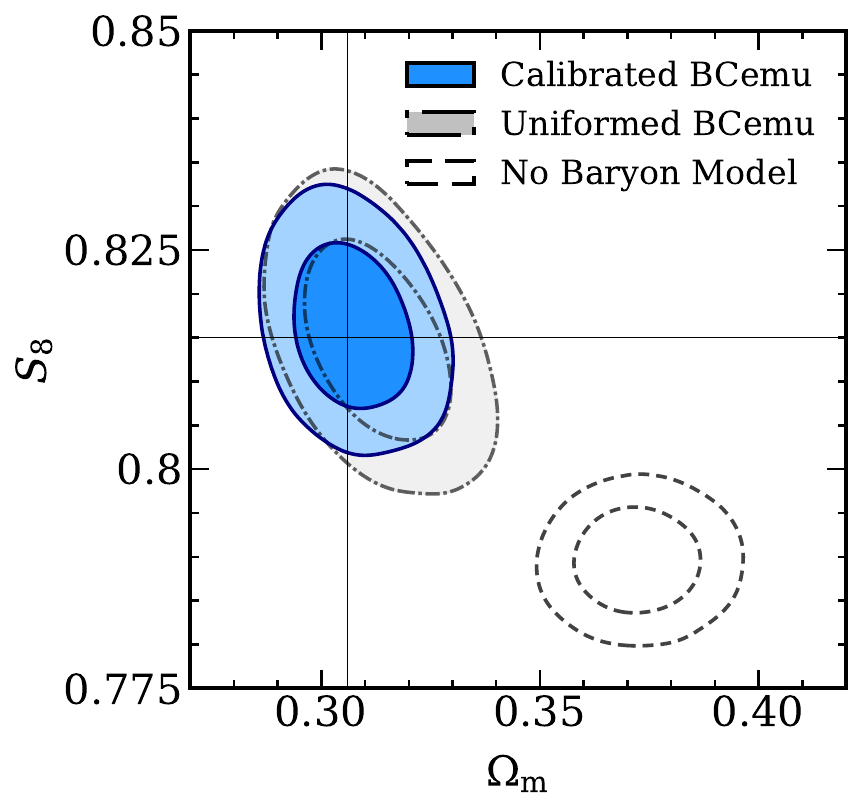}

\caption{
Simulated constraints on the matter density, $\Omega_{\rm m}$, and the lensing amplitude, $S_8$, for an LSST Year~1 cosmic shear analysis \citep{lsst_srd}, demonstrating the improvement in precision and accuracy afforded by our calibrated model. 
Without a baryon feedback model, the inferred cosmology is significantly biased (black, dashed);
this constraint is unphysical but indicative of the available constraining power if feedback was precisely known.
Due to its added flexibility, the uniformed baryonification model (black, dot-dashed) degrades the constraining power by $2\times$, relative to the inference without a baryon model.  
Our kSZ and eROSITA  calibrated baryonification prior (blue) recovers half the precision otherwise lost to the flexible baryon model. 
}
\label{fig:LSST}
\end{figure}

Modelling the impact of baryon feedback on the matter power spectrum is a limiting systematic for weak lensing surveys. Here we assess how our feedback constraints on the baryonification model parameters improve weak lensing cosmological precision and accuracy. We consider an LSST Year~1-like analysis \citep{lsst_srd}, comparing results with and without our calibrated model priors from current kSZ and eROSITA data.

To generate mock cosmic shear data, we follow the LSST Dark Energy Science Collaboration Science Requirements Document \citep{lsst_srd}. We adopt their specifications for source density, redshift binning, redshift distributions, and photometric redshift uncertainties. The data vector is constructed using the COSMOSIS framework \citep{zuntz2015} as described in \citet{Preston2024}. 
The nonlinear matter power spectrum is treated with HMCode \citep{Mead2021}. We assume the best-fit baryonification parameters from the joint kSZ and eROSITA fit as the input level of feedback.

In Figure~\ref{fig:LSST}, we report the constraints on $\Omega_{\rm m}$ and $S_8$ for three different treatments of baryon feedback: i) no baryon model, ii) baryonification with uniform wide priors, and iii) baryonification with kSZ and eROSITA calibrated priors. 
As expected, the inferred cosmology is significantly biased without a baryon model.
Although unphysical, these constraints illustrate the expected cosmological precision if baryon feedback was perfectly known. 
The uninformed baryonification model recovers the input cosmology, however, the added model flexibility degrades the cosmological constraining power by $2\times$ relative to the inference without a baryon model. 
The kSZ and eROSITA calibrated baryonification model accurately constrains $\Omega_{\rm m}$ and $S_8$ and recovers half the precision otherwise lost to the flexible baryon model.
With data-driven constraints on baryon feedback, early LSST analyses can robustly exploit small-scale information. 

\section{Conclusions}
\label{sec:conc}
We have presented the most comprehensive multi-wavelength constraints to date on how baryonic feedback suppresses the non-linear matter power spectrum, a key uncertainty for cosmic shear cosmology. 
By jointly analysing SDSS/DESI+ACT kSZ effect profiles \citep{Schaan2021,Ried2025} and X-ray gas mass fractions from eROSITA \citep{Bulbul2024} and HSC--XXL \citep{Pierre2016,Akino2022}, with weak lensing halo masses \citep{Siegel2025flamingo}, we directly probe the distribution of gas in groups and clusters ($13<\log_{10} \frac{M_{500}}{ \mathrm{M}_\odot~h^{-1}} <14$) at $0<z<1$.
Using the baryonification framework \citep{Schneider2015,Giri2021}, our joint constraints reveal that baryons significantly reshape the matter power spectrum  on small scales.

Our main findings are:
\begin{itemize}
\item The kSZ effect profiles and eROSITA hot gas mass fractions independently and consistently favour strong feedback, implying a 
$10 \pm 2\%$ suppression of the matter power spectrum at $k=1~h~\mathrm{Mpc}^{-1}$. This suppression exceeds the predictions of the fiducial models from recent hydrodynamical simulation suites, e.g. FLAMINGO and BAHAMAS, which were calibrated to reproduce pre-eROSITA X-ray measurements, including the HSC--XXL sample. 
We demonstrate the robustness of our results to observational systematics and perform successful injection recovery tests of the model's ability to map between gas observables and matter power suppression.

\item The HSC–XXL X-ray scaling relation favours more moderate feedback than the kSZ and eROSITA data, corresponding to only $5 \pm 4\%$  suppression at $k=1~h~\mathrm{Mpc}^{-1}$. 
Potential explanations include residual X-ray calibration differences, X-ray selection-function effects, X-ray modelling uncertainties, and/or unaccounted-for systematics in the kSZ velocity reconstruction. 
Cross-instrument calibration and forward modelling efforts will be critical to understanding this discrepancy. 
Nevertheless, all datasets consistently support non-zero suppression of the matter power spectrum, reinforcing the conclusion that constraining baryonic feedback is essential to modelling small-scale structure. 

\item For upcoming surveys, such as LSST, Euclid, and Roman,  observationally calibrated feedback models offer a path to unlocking small-scale cosmic shear. With a simulated LSST Year 1 analysis, we show that this framework can allow weak lensing analyses to safely extend to smaller angular scales, reducing the cosmological degradation associated with conservative scale cuts or flexible feedback models.

\end{itemize}

Understanding baryon feedback is not only an astrophysical problem but also a direct limitation on weak lensing cosmological constraining power.
External constraints on the gas distribution are therefore an indispensable ingredient for next-generation analyses.
X-ray and kSZ measurements will improve dramatically with upcoming data from eROSITA, Simons Observatory, and DESI \citep[e.g.][]{Ried2025, Wayland2025, Lague2025}, requiring greater understanding of systematic uncertainties.
Forward modelling studies of X-ray gas mass fractions and kSZ effect profiles will be crucial to understanding observational and modelling systematics.
In the near future, additional measurements of the gas distribution will add complementary information to constrain baryon feedback \citep[e.g.][]{Dalal2025, Pandey2025, Sharma2025}. 
Feedback models that are calibrated with these upcoming multi-wavelength observables can realize the small-scale statistical power of cosmic shear.

\section*{Acknowledgements}
We thank George Efstathiou for detailed feedback on this manuscript.
JS acknowledges support by the National Science Foundation Graduate Research Fellowship Program under Grant DGE-2039656. 
LB acknowledges support from the Science and Technology Facilities Council (STFC). 
Any opinions, findings, and conclusions or recommendations expressed in this material are those of the author(s) and do not necessarily reflect the views of the National Science Foundation. 
This work was supported by the Science and Technology Facilities Council (grant number ST/Y002733/1).

%%%%%%%%%%%%%%%%%%%%%%%%%%%%%%%%%%%%%%%%%%%%%%%%%%
\section*{Data Availability}

The data underlying this article may be shared on reasonable request
to the corresponding authors.



\bibliographystyle{mnras}
\bibliography{references} % if your bibtex file is called example.bib

%%%%%%%%%%%%%%%%%%%%%%%%%%%%%%%%%%%%%%%%%%%%%%%%%%

%%%%%%%%%%%%%%%%% APPENDICES %%%%%%%%%%%%%%%%%%%%%

\appendix

\section{Sensitivity to Modelling Choices}
\label{app:sensitivity}

\begin{figure*}
\centering
\includegraphics[width=1.15\columnwidth]{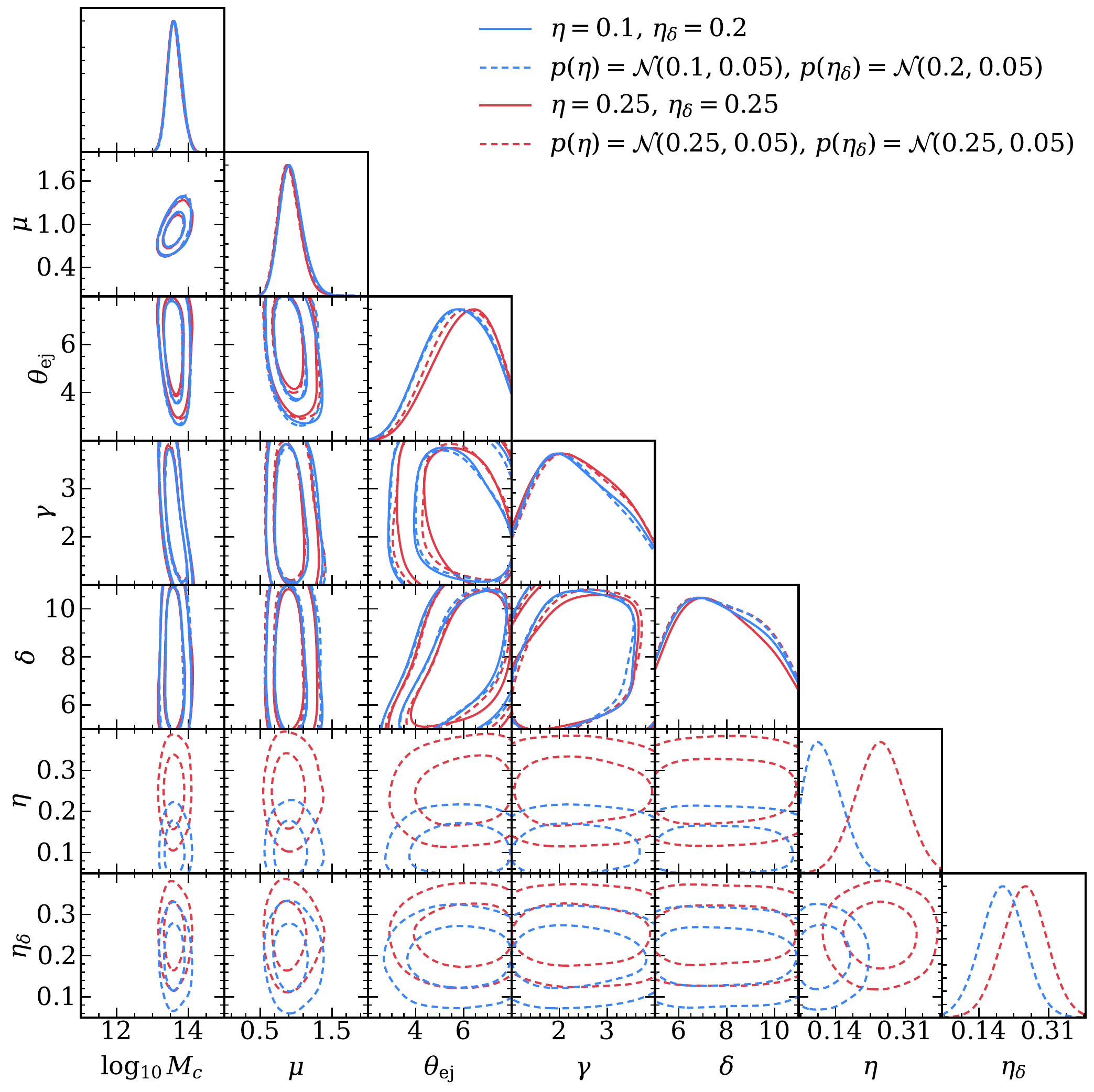}
\includegraphics[width=0.85\columnwidth]{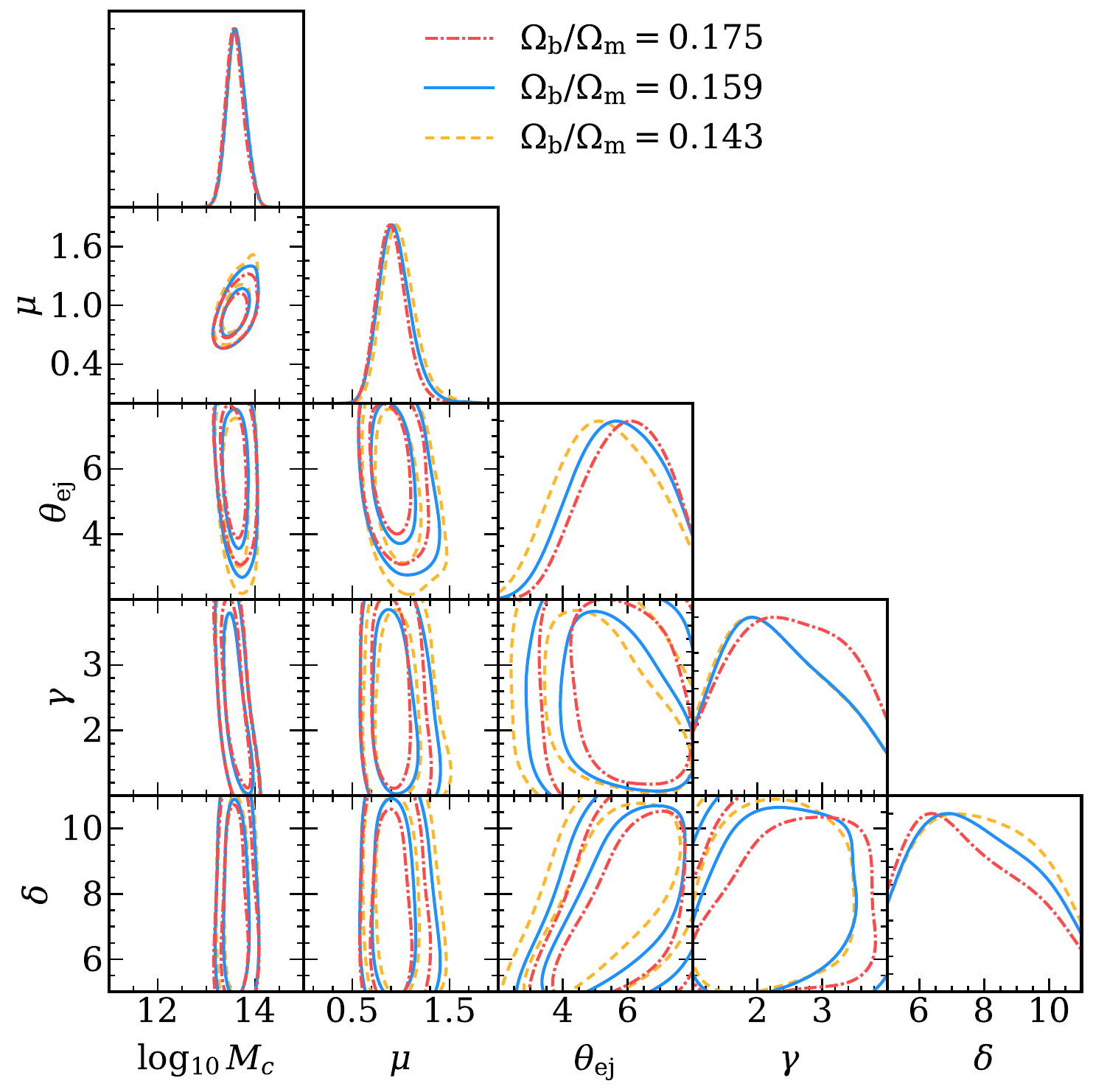}
    \caption{
    The posterior constraints on the baryonification parameters are weakly sensitive to the assumed stellar parameters ($\eta$, $\eta_\delta$) and universal baryon fraction ($\Omega_\mathrm{b} / \Omega_\mathrm{m}$).
    \textit{Left:} 
    the joint eROSITA X-ray and kSZ parameter constraints for four choices of stellar parameter priors. 
    Our fiducial fit assumes $\eta=0.1$ and $\eta_\delta=0.2$ (solid blue).
    The dashed blue contours assume conservative Gaussian priors:  $p(\eta)=\mathcal{N}(0.1,0.05)$ and $p(\eta_\delta)=\mathcal{N}(0.2,0.05)$.
    To test the sensitivity of the posteriors to the adopted prior, we also consider $\eta=0.25$ and $\eta_\delta=0.25$ (solid red); the dashed contours again correspond to Gaussian priors:  $p(\eta)=\mathcal{N}(0.25,0.05)$ and $p(\eta_\delta)=\mathcal{N}(0.25,0.05)$.
    \textit{Right:} 
    the posterior constraints for three choices of $\Omega_\mathrm{b} / \Omega_\mathrm{m}$.
    We present our fiducial choice in the solid contour \citep[the maximum likelihood of the  DES~Y3 `3×2pt + All Ext.' flat $\Lambda$CDM cosmology,][]{DES3x2} and  show the results for $10\%$ lower and higher $\Omega_\mathrm{b} / \Omega_\mathrm{m}$ in dashed and dot-dashed contours, respectively.
    }
    \label{fig:corner_piors}
\end{figure*}

In this work, we constrain the suppression of the matter power spectrum, {\pk}, by modelling kSZ effect profiles and X-ray gas mass fractions with the baryonification framework \citep{Schneider2015}.
Because stellar fraction measurements are currently unavailable for our eROSITA cluster sample, we set the stellar parameters to $\eta=0.1$ and $\eta_\delta=0.2$ for our fiducial analysis following \cite{Kovac25}.
We also fix the universal baryon fraction, $\Omega_\mathrm{b} / \Omega_\mathrm{m}$, to the maximum likelihood cosmological parameters of the DES~Y3 `3×2pt + All Ext.' flat $\Lambda$CDM cosmology \citep{DES3x2}.
In this appendix, we investigate the sensitivity of our results to these choices. 

We repeat the joint kSZ and eROSITA gas mass fraction fit using four different stellar parameter priors:
i) $\eta=0.1$ and $\eta_\delta=0.2$ (fiducial), ii) $p(\eta)=\mathcal{N}(0.1,0.05)$ and $p(\eta_\delta)=\mathcal{N}(0.2,0.05)$,  iii)  $\eta=0.25$ and $\eta_\delta=0.25$ \citep[motivated by][]{Grandis2024}, and iv) $p(\eta)=\mathcal{N}(0.25,0.05)$ and $p(\eta_\delta)=\mathcal{N}(0.25,0.05)$.
We also repeat the joint kSZ and X-ray fit for two alternate baryon fractions: $10\%$ lower and higher than the DES~Y3 `3×2pt + All Ext.' flat $\Lambda$CDM maximum likelihood; assuming our fiducial stellar parameter choice ($\eta=0.1$ and $\eta_\delta=0.2$).
The posteriors of each fit are presented in Figure~\ref{fig:corner_piors}. 
The different stellar parameter and baryon fraction assumptions impact the posterior constraints below the $1\sigma$ level; this is consistent with prior baryonification modelling results \citep[e.g.,][]{Giri2021,Grandis2024}.

\section{Consistency of the kSZ and X-ray data}
\label{app:xray}

\begin{figure*}
\centering
\includegraphics[width=\columnwidth]{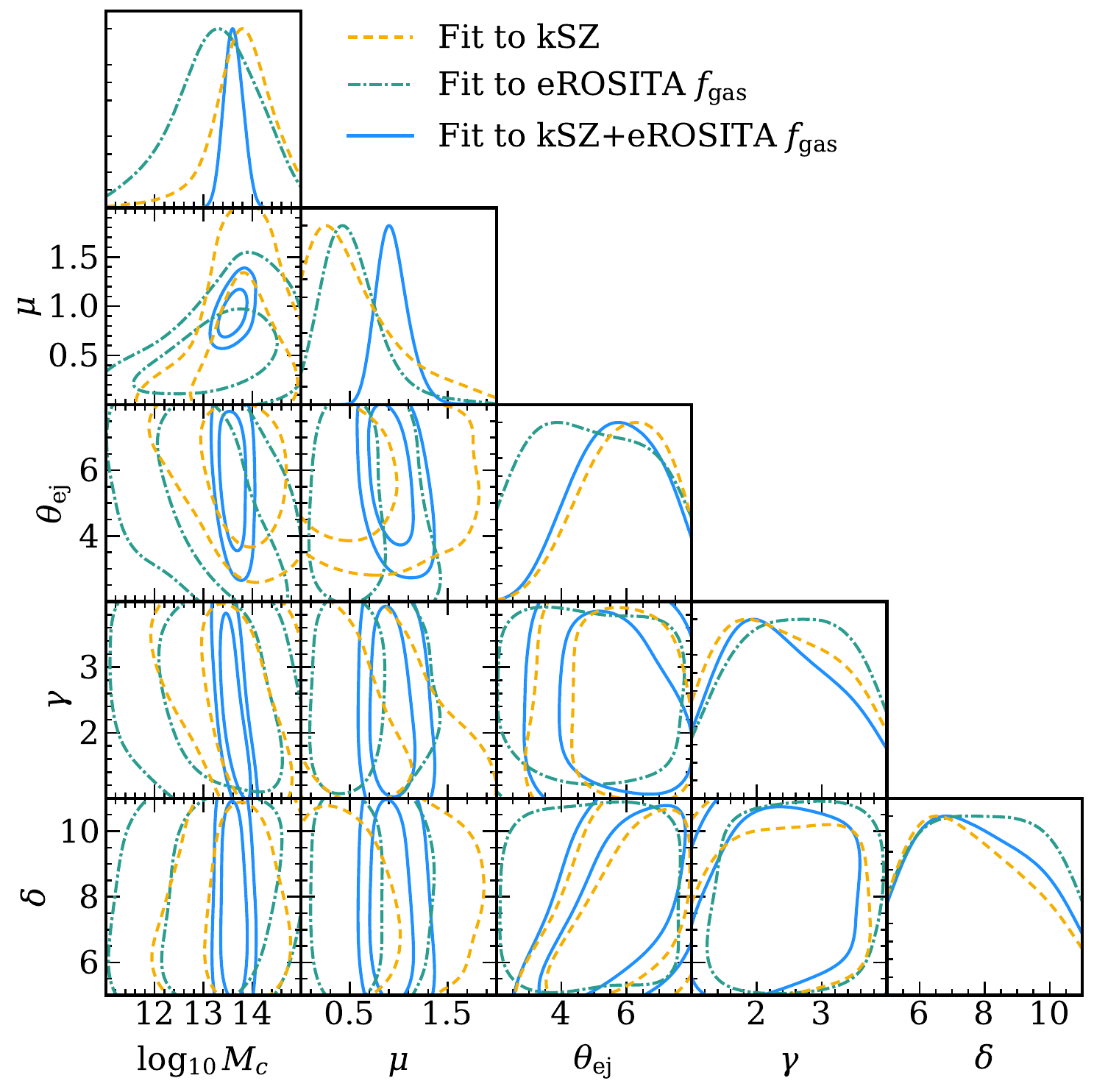} \includegraphics[width=\columnwidth]{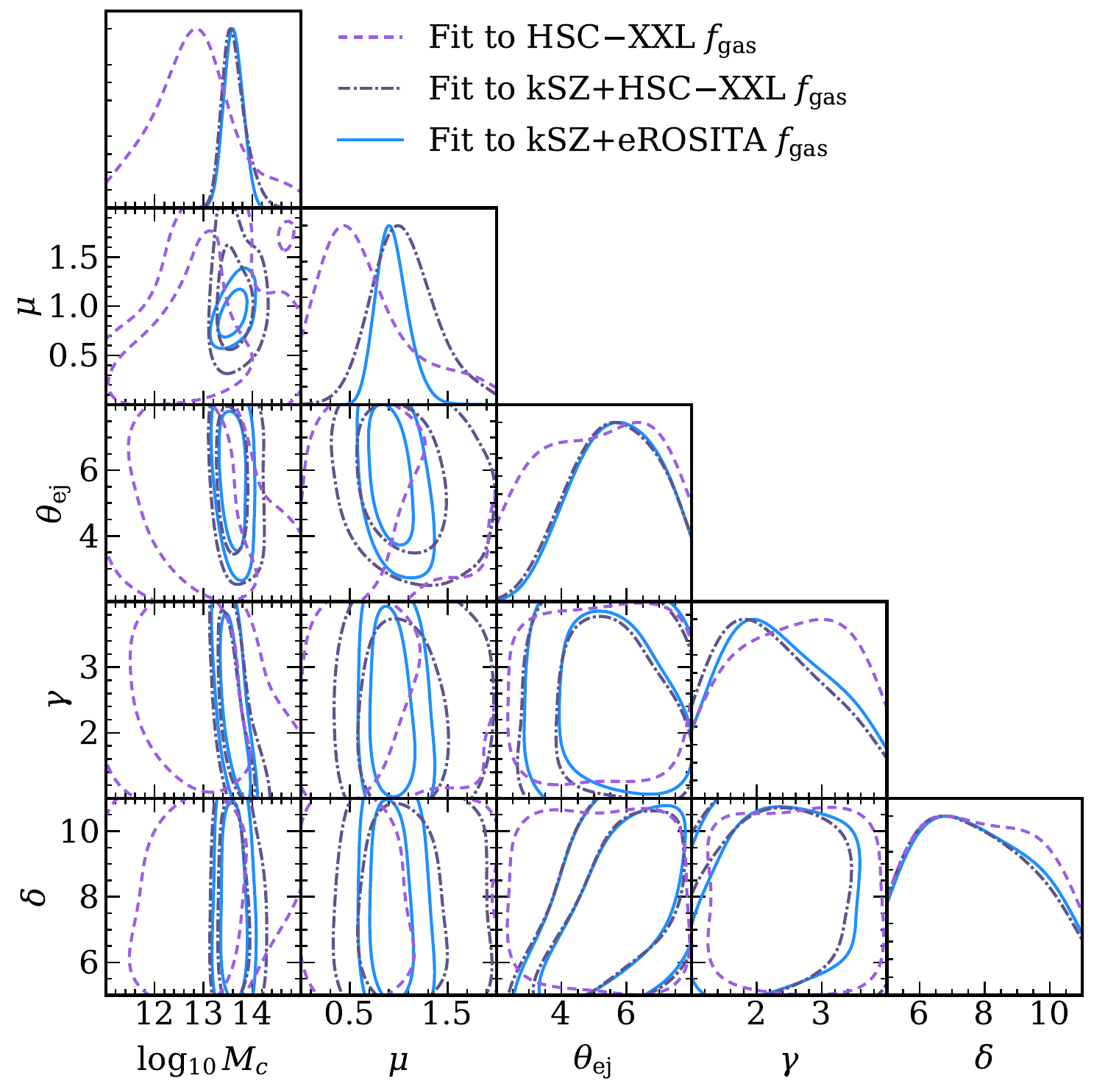}
    \caption{Posteriors on the baryonification parameters from fitting the kSZ and X-ray data.
    \textit{Left:} posteriors for the joint kSZ and eROSITA fit (solid), alongside the fits to the kSZ and eROSITA data separately (dashed and dot-dashed, respectively). 
    \textit{Right:} posteriors from the HSC--XXL gas mass fractions alone (dashed) and jointly with kSZ (dot-dashed).
    For comparison, the joint kSZ and eROSITA posteriors are again shown in solid-blue.
    }
    \label{fig:corner_data}
\end{figure*}

To constrain the baryonification model, we consider SDSS/DESI+ACT kSZ effect profiles \citep{Schaan2021,Ried2025} and X-ray gas mass fraction measurements from the eROSITA \citep{Bulbul2024} and HSC--XXL samples \citep{Pierre2016,Akino2022}.
In Section~\ref{sec:consistency}, we compare the posterior constraints from fitting the kSZ, eROSITA, and HSC--XXL data separately.
The posteriors of each fit are presented in Figure~\ref{fig:corner_data}.
Consistent with previous studies, we find a slight tension between the strength of feedback implied by the kSZ effect profiles and pre-eROSITA X-ray gas mass fractions \citep{bigwood2024,McCarthy2025,Kovac25}.

\begin{figure*}
\centering
\includegraphics[width=0.8\textwidth]{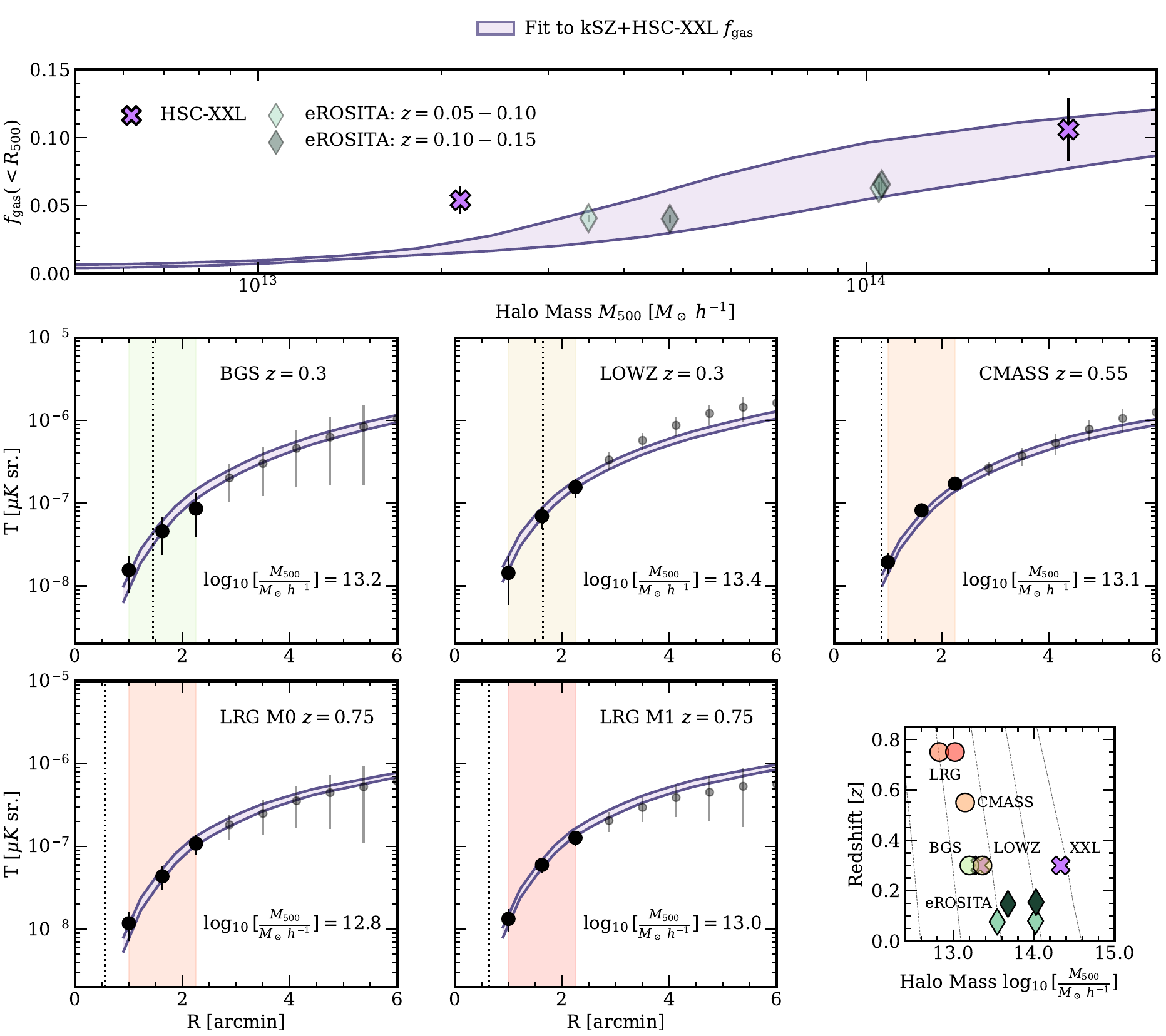}
    \caption{
    The baryonification model jointly fit to the kSZ and HSC--XXL X-ray measurements (purple), alongside the observations.
    Refer to Figure~\ref{fig:goodness_of_fit_mega} for a detailed description.
    }
    \label{fig:goodness_of_fit_mega_akino}
\end{figure*}

Although the baryonification model provides an excellent joint fit to the kSZ and eROSITA measurements (Figure~\ref{fig:goodness_of_fit_mega}), it struggles to simultaneously describe the kSZ and HSC--XXL data.
Figure~\ref{fig:goodness_of_fit_mega_akino} compares the joint kSZ and HSC--XXL fit with the data.
The poor goodness of fit is primarily driven by the lower HSC--XXL mass bin and the BGS and LOWZ kSZ profiles;
these measurements probe similar radii, redshift, and halo mass, but appear to require different levels of gas depletion.
We discuss possible explanations for this discrepancy in Section~\ref{sec:xray_disc}.

\section{Injection recovery}\label{app:injrec}

The baryonification framework relates gas observables (e.g., X-ray gas mass fractions) to the suppression of the matter power spectrum \citep{Schneider2015,Giri2021,Schneider2025}.
Using the BAHAMAS, OWLS, Illustris, and Horizon simulations, \cite{Giri2021} validated the \texttt{BCemu} baryonification emulator by accurately predicting the suppression of the matter power spectrum from the gas and stellar mass fractions of simulated groups and clusters and vice versa. 
In this work, we further validate the framework using the FLAMINGO simulations (see Section~\ref{sec:inj}); for brevity, we consider the fiducial (L1\_m9) and strongest ({\eightsigma}) feedback variants of the FLAMINGO suite.

\begin{figure*}
\centering
\includegraphics[width=0.85\textwidth]{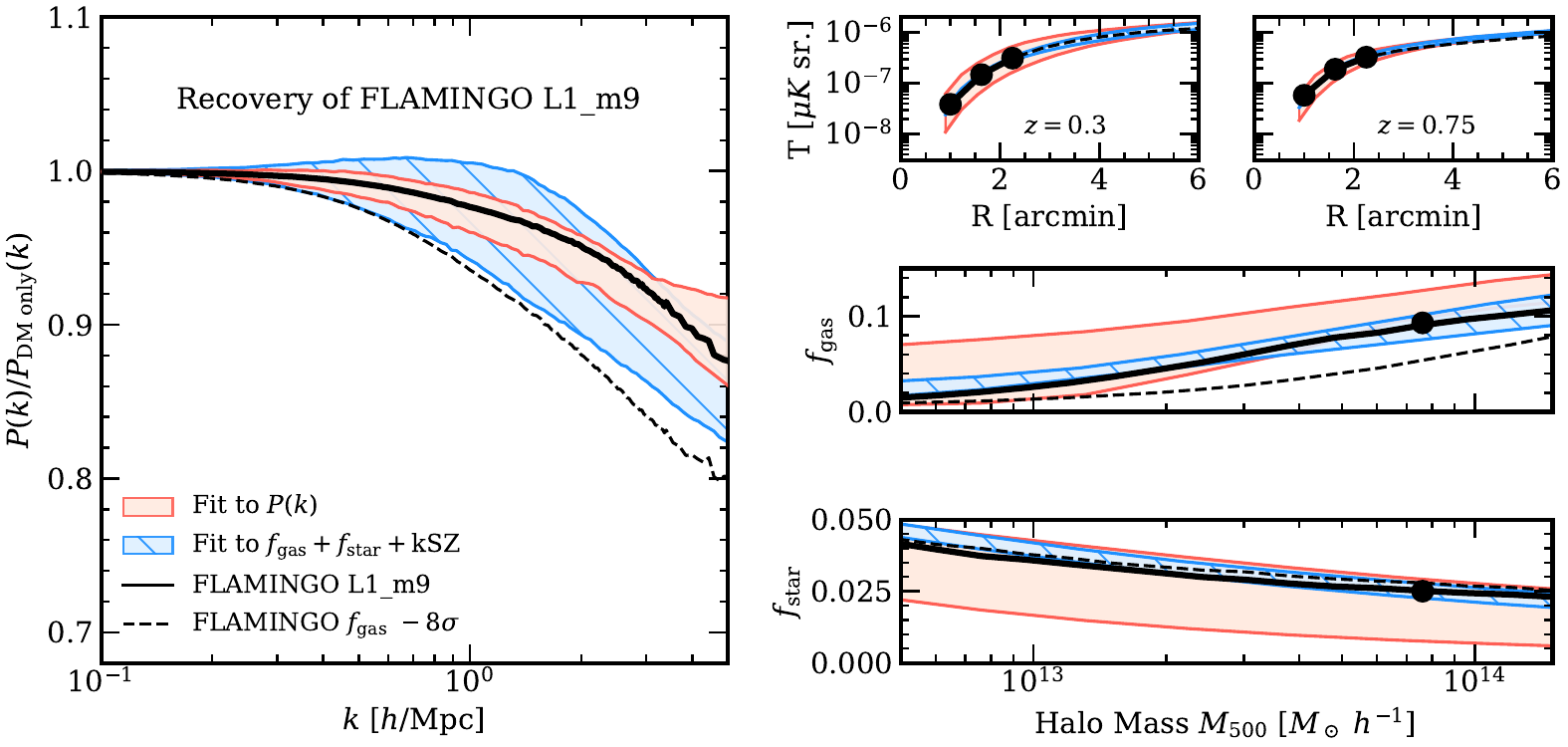}
\includegraphics[width=0.85\textwidth]{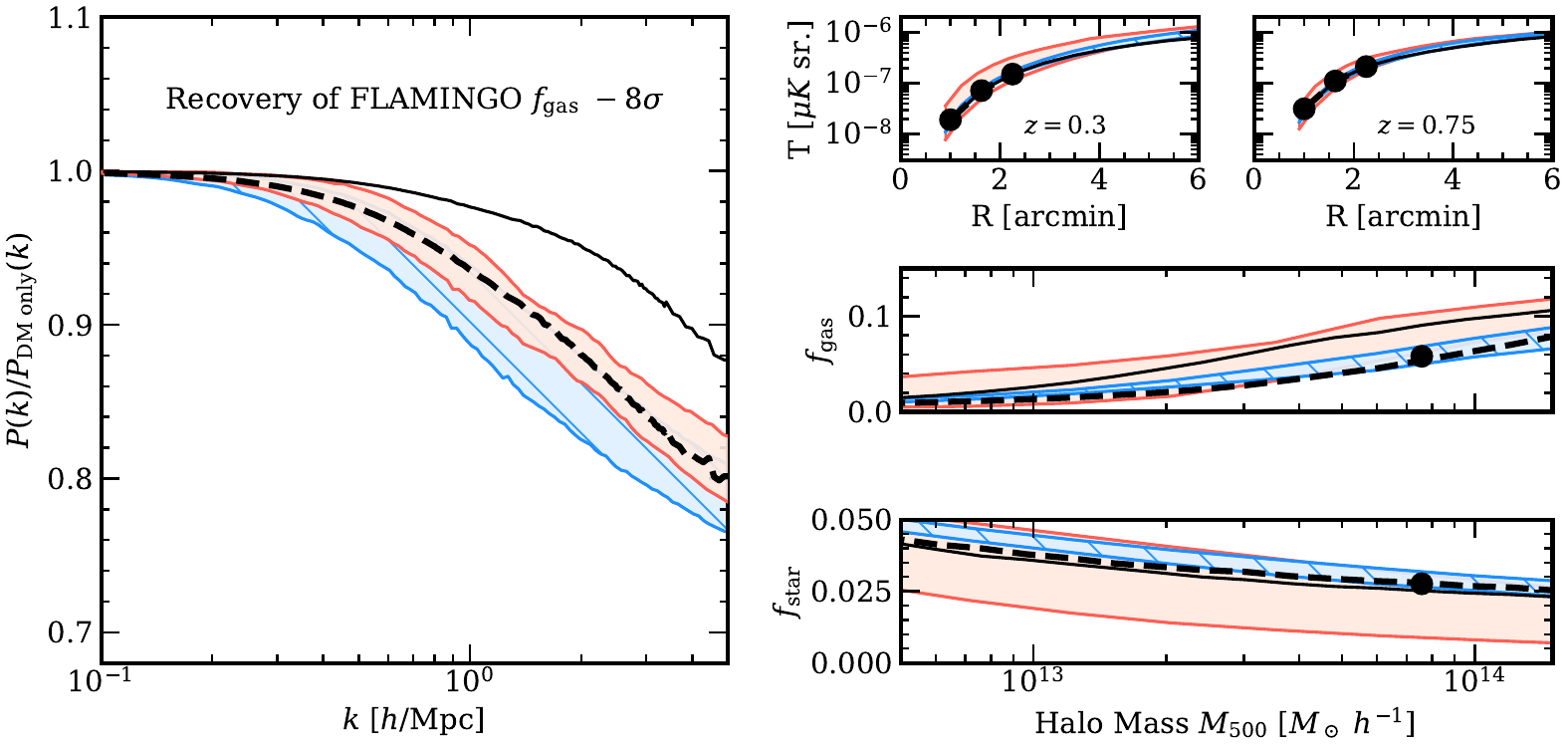}
\caption{
The baryonification model successfully predicts the suppression of the matter power spectrum {\pk} from mock kSZ, gas mass fraction, and stellar mass fraction observations, and vice versa.
Refer to Figure~\ref{fig:injrecovery1} for a detailed description.
    }
    \label{fig:injection_recovery_with_fstar}
\end{figure*}

\begin{figure*}
\centering
\includegraphics[width=\textwidth]{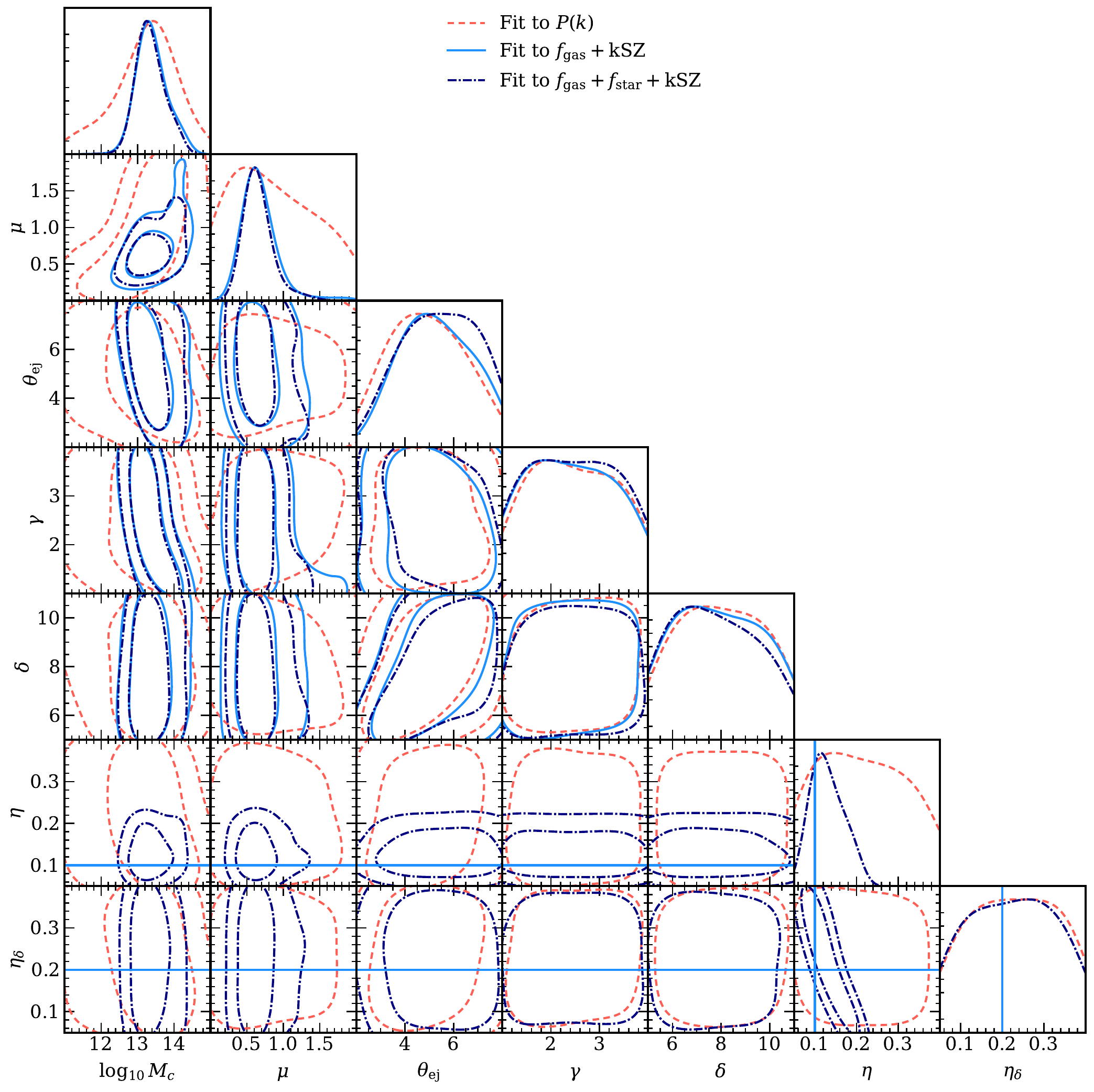}
    \caption{Posteriors on the baryonification parameters from the FLAMINGO simulation injection--recovery tests.
    The constraints from fitting the matter power suppression, {\pk}, are shown in pink (dashed).
    The joint kSZ and X-ray gas mass fraction fit is shown in blue (solid);
    following our analysis of the data, we fix the stellar parameters to $\eta=0.1$ and $\eta_\delta=0.2$.
    The joint kSZ, gas mass fraction, and stellar mass fraction fit is shown in purple (dot-dashed).
    For brevity, we present the results from the strongest feedback FLAMINGO variant ({\eightsigma}), because it matches the SDSS/DESI+ACT kSZ data best \citep{McCarthy2025,Siegel2025flamingo}.}
    \label{fig:corner_mock}
\end{figure*}

The model accurately predicts the gas and stellar content of groups and clusters from {\pk}, for the fiducial and strongest feedback FLAMINGO simulations.
The framework also successfully predicts  {\pk} from group and cluster observables.
To reflect our data fits, in Section~\ref{sec:inj} we considered the stacked kSZ profiles of group mass halos at $z=0.3$ and $0.75$ and the gas mass fractions of clusters.
For completeness, here we consider a joint kSZ, gas mass fraction $f_\mathrm{gas}$, and stellar mass fraction $f_\mathrm{star}$ fit with uniform priors for all 7 baryonification parameters
(Figure~\ref{fig:injection_recovery_with_fstar}).
Both the kSZ$+f_\mathrm{gas}$ fit and the kSZ$+f_\mathrm{gas}+f_\mathrm{star}$ fit accurately predict the matter power suppression.
The posteriors of the {\pk}, kSZ$+f_\mathrm{gas}$, and kSZ$+f_\mathrm{gas}+f_\mathrm{star}$ fits are presented in Figure~\ref{fig:corner_mock}.
The recovered parameters are consistent between all three fits.

\section{Challenges of kSZ modelling}
\label{app:two_halo}

The temperature change induced by the kSZ effect is proportional to the density and velocity of the gas along the line of sight  (Equation~\ref{eq:ksz1}). 
For an idealized halo---spherically symmetric gas distribution moving at the same velocity as the central---the conversion between the gas density profile and the kSZ effect profile is well-defined. 
The presence of satellite galaxies and two-halo interactions complicates this picture.

\begin{figure*}
\centering
\includegraphics[width=\textwidth]{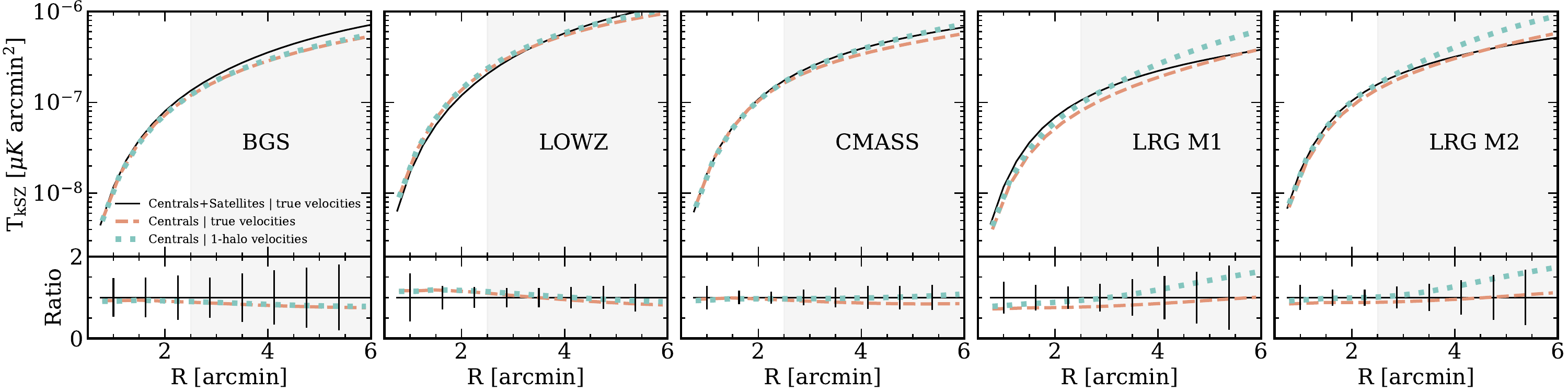}
    \caption{Demonstration of how  satellites and two-halo interactions alter the kSZ effect profile in FLAMINGO;
    for brevity, we present the results from the strongest feedback FLAMINGO variant ({\eightsigma}), because it matches the SDSS/DESI+ACT kSZ data best.
    We present the simulated kSZ effect profile for samples with (solid) and without (dashed) satellites.
    The simulated galaxies are selected by matching the measured GGL profiles of each SDSS/DESI+ACT sample \citep{Siegel2025flamingo}. 
    We also test the approximation that the peculiar velocity of the gas distribution is the same as the velocity of the central galaxy.
    For the centrals only samples, the dotted lines present the kSZ profile where the gas particle velocities are fixed to the central halo velocity.
    Each column corresponds to a different SDSS/DESI+ACT measurement, and the bottom row presents the ratio of the different kSZ signals with the fiducial calculation.
    The measurement uncertainties are shown alongside the residuals for context.
    The shaded region ($>3'$) demarcates the scales omitted from our analysis due to modelling uncertainties. 
    }
    \label{fig:modeling_uncertainties}
\end{figure*}

Satellites potentially impact the measurement of the kSZ effect in several ways \citep{McCarthy2025}.
Satellites will be miscentered from the centrals, thereby lowering the kSZ signal
relative to a galaxy at the centre of the gas distribution.
Satellites also occupy more massive halos than expected from stellar mass alone;
because halo mass and gas mass are approximately proportional, satellites potentially boost the kSZ signal relative to a sample of all centrals.
For a fixed stellar mass selection (i.e., all galaxies above a minimum stellar mass cut), \cite{McCarthy2025} found that the kSZ effect signal from a sample of centrals is lower than with satellites. 
The effect is approximately $30\%$ for a low mass selection ($\log_{10} M_\star / \mathrm{M}_\odot > 11$) and only $5\%$ for $\log_{10} M_\star / \mathrm{M}_\odot > 11.6$.
This effect is partially a result of the different mean halo masses between the samples with and without satellites.
For each SDSS/DESI+ACT kSZ stack, \cite{Siegel2025flamingo} identified the
selection of simulated galaxies that best fits the measured GGL signal. 
To isolate the impact of satellites, we compare the mock kSZ effect signals for the best-fit selection of centrals and the best-fit selection of centrals and satellites in  Figure~\ref{fig:modeling_uncertainties}.
For small angular scales ($<3'$), the two selections agree well.
On larger scales, the kSZ effect signal from the centrals only selection is lower than the sample with satellites, potentially reflecting differences in the underlying gas distribution.

To predict the kSZ effect from a given gas density profile, we assume that the gas distribution has the peculiar velocity of the central galaxy. 
This assumption is reasonable on small scales, but on larger scales the gas velocity is not necessarily dominated by a single central potential.
To test this approximation, we calculate the kSZ effect profile from FLAMINGO using either the true per-particle velocities or assigning the gas particles the peculiar velocity of the central host. 
We compare the true and approximate kSZ effect profiles in Figure~\ref{fig:modeling_uncertainties}; the kSZ profiles are calculated for the selection of central halos that best fit the measured GGL signal for each SDSS/DESI+ACT kSZ stack.
As expected, the true and approximate kSZ effect profiles are identical on small scales, where it is reasonable to assume that the gas velocity is the same as the central galaxy velocity, but the approximation breaks down on larger scales.
The difference is most pronounced at high redshift ($z=0.75$), because the kSZ profile is probing beyond several $R_{500}$.

For a given sample of galaxies, the kSZ effect is typically measured by a weighted stack of the CMB temperature map \citep{Schaan2021,Hadzhiyska2024photoz,Ried2025}, where the weights are the galaxies' reconstructed peculiar velocities.   
The fidelity of the velocity reconstruction is a potential source of uncertainty. 
To account for non-linearities and shot noise uncertainties, current studies calculate a correction factor based on mock catalogue velocity reconstructions \citep{Schaan2021,Hadzhiyska2024photoz};
for the DESI samples, the correlation between the true and reconstructed velocities is approximately $0.7$ with quoted uncertainties of less than $5\%$ \citep{Hadzhiyska2024}. 
For each SDSS/DESI+ACT kSZ stack, the selection of simulated galaxies that best fits the GGL signal has a peculiar velocity RMS within $10\%$ of the bias corrected linear theory reconstruction \citep{Siegel2025flamingo}.
Systematics could be further investigated by reconstructing the simulated galaxy velocities using the same method as the observations.

To mitigate uncertainties, particularly the effect of satellites and two-halo interactions, we restrict our kSZ analysis to small scales $<3'$.
Because the kSZ effect measurement is highly correlated at larger angular separations \citep{Ried2025}, these scale cuts do not significantly reduce the signal-to-noise ratio of the data. 
Future work addressing these uncertainties is critical to unlocking the full potential of the SDSS/DESI+ACT measurements. 

% Don't change these lines
\bsp	% typesetting comment
\label{lastpage}
\end{document}